\documentclass[twocolumn,10pt]{revtex4}%
\usepackage[paperwidth=210mm,paperheight=297mm,centering,hmargin=2cm,vmargin=2.5cm]{geometry}
\usepackage{amsfonts}
\usepackage{amsmath}
\usepackage{amssymb}
\usepackage{bm}
\usepackage{graphicx}
\usepackage{color}
\usepackage{soul}
\usepackage{epsfig}%
\usepackage[dvipsnames]{xcolor} 
  \definecolor{bleu_cite}{RGB}{0,0,255}
\usepackage[colorlinks=true,allcolors = black,citecolor=bleu_cite,  ]{hyperref} 
\usepackage{natbib}
\usepackage{siunitx}

\def\fd{\textcolor{black}}

\begin{document}
\title{Mott insulator of strongly interacting two-dimensional semiconductor excitons}

\author{Camille Lagoin$^1$, Stephan Suffit$^{2}$, Kirk  Baldwin$^3$, Loren Pfeiffer$^3$ and Fran\c{c}ois Dubin$^{1}$} 
\affiliation{$^1$ Institut des Nanosciences de Paris, CNRS and Sorbonne Universit{\'e}, 4 pl. Jussieu,
75005 Paris, France}
\affiliation{$^2$ Laboratoire de Materiaux et Phenomenes Quantiques, Universit{\'e} Paris Diderot, 75013 Paris, France}
\affiliation{$^3$ PRISM, Princeton Institute for the Science and Technology of Materials, Princeton University, Princeton, NJ 08540, USA}

\begin{abstract}
\textbf{In condensed-matter physics, electronic Mott insulators have triggered considerable research due to their intricate relation with high-temperature superconductors. However, unlike atomic systems for which Mott phases were recently shown for both bosonic and fermionic species, in the solid-state the fingerprint of a Mott insulator implemented with bosons is yet to be found. Here we unveil such signature by exploring the Bose-Hubbard hamiltonian using semiconductor excitons confined in two-dimensional lattices. We emphasise the regime where on-site interactions are comparable to the energy separation between  lattice confined states. We then observe that Mott phases are accessible, with at most two excitons uniformly filling lattice sites. The technology introduced here allows us to program on-demand the geometry of the lattice confining excitons. This versatility, combined with the long-range nature of dipolar interactions between excitons, provide a new route to explore many-body phases \fd{spontaneously breaking the lattice symmetry.}}
\end{abstract}

\maketitle

For strongly correlated many-body systems, the interplay between the interaction strength and the kinetic energy can lead to intriguing quantum phases unexpected at first. This is notably the case when  fermions/bosons are confined in a lattice potential, since an incompressible state with the same integer number of particles per lattice site becomes energetically preferred above a critical interaction strength \cite{Salomon_2010}. Such Mott insulator (MI) has received a considerable attention, for electrons in a wide class of materials \cite{Book_Mott,Mott_supra,graphene_bilayer,Wign_1,Wign_2,Wign_3}, and more recently for both bosonic and fermionic ultra-cold atoms \cite{Greiner_2002,Chin_2009,Greiner_Mott,Bloch_2010,Esslinger_fermions,Bloch_fermions,Greiner_fermions,Ye_2017}. Indeed, a Mott phase provides the building block to study strongly correlated quantum many-body states such as high-temperature superconductivity \cite{Book_Mott} or lattice supersolidity \cite{Maciej_2002,Dutta_2015,Baranov_2011,Trefzger_2010}. 

For bosonic systems, a MI is specifically captured by the celebrated Bose-Hubbard (BH) hamiltonian \cite{Salomon_2010}. In its simplest form this model is restricted to a single state per lattice site, i.e. a single Wannier state (WS), and includes the strength of on-site interaction $U$ together with the tunnelling strength between nearest neighbouring sites $t$ (Fig.1.a). Over the last two decades studies of ultra-cold atomic vapours have deeply scrutinised the  physics of the BH model, e.g. evidencing the transition between superfluid and Mott phases \cite{Greiner_2002,Chin_2009,Greiner_Mott,Bloch_2010}. Strong research efforts are now directed towards extending the BH hamiltonian to interactions between adjacent lattice sites  \cite{Ferlaino_2016}. Indeed, this is necessary to realise many-body phases spontaneously breaking the lattice symmetry, such as stripes, checkerboards or lattice supersolids \cite{Maciej_2002,Dutta_2015,Baranov_2011,Trefzger_2010}. Extending the BH model to the multi-orbital regime, where a discrete spectrum of WS is accessible, constitutes another challenging direction \cite{Dutta_2015,MC_BH,Mark_2011} particularly relevant in the solid-state, notably for moiré excitons in atomically thin monolayers \cite{moire_1,moire_2,moire_3,moire_4}.

Here, we explore the phase diagram of the BH model in a multi-orbital regime controlled by repulsive dipolar interactions. For that we rely on two-dimensional semiconductor excitons confined in square lattice potentials with  several hundreds nanometre periods. Excitons are then subject to on-site interactions with a magnitude $U$  comparable to the energy separation between successive WS, \fd{$\Delta E$}. We then observe that MI phases are accessible \fd{in the lowest energy WS for which $U$ does not exceed $\Delta E$}.

Figure 1.b illustrates the semiconductor technology that we have developed. It is based on two adjacent GaAs quantum wells, each confining electrons or holes that are spatially separated but Coulomb bound to realise dipolar excitons \cite{Combescot_ROPP} (Fig.S4.a). These are non resonantly injected using a pulsed laser excitation (Fig.S4), repeated at 850 kHz. The excitation has a gaussian spatial profile, its average power $P$ controling the mean density in the lattice. In the following, excitons are studied at a bath temperature T=330 mK, in a 50 ns long time interval set 250 ns after extinction of the laser pulse. This delay is about one third of the excitons optical lifetime (Fig.S6). Note that for every experimental settings we statistically average 10 measurements performed under the same conditions, each measurement lasting typically 30 seconds thus averaging $25\cdot 10^6$ realisations.

Using nano-patterned metallic electrodes deposited  at the surface of the field-effect structure embedding the GaAs bilayer (Fig.1.b and Fig.S1), we engineer spatially homogeneous square lattice potentials \cite{Lagoin_2020,Lagoin_2021}. For that, we exploit the interaction between the excitons large permanent electric dipole and the spatially varying electric field  defined by the gate electrodes \cite{High_2009,Grosso_2009,Winbow_2011,Shilo_2013,Schinner_2013}. In the following, we report two devices implementing 400 and 800 nm period lattices, both having a depth set to around 1.5 meV (section I of Supplementary Informations (SI)). Accordingly, in the 800 nm period lattice 15 WS are accessible \fd{with $\Delta E\sim$ 100 $\mu$eV} (Fig.1.c), whereas the spectrum reduces to 8 WS for the 400 nm period device \fd{with $\Delta E\sim$ 200 $\mu$eV} (Fig.S8).

\fd{To quantify our experimental findings we evaluated the parameters $U$ and $t$ of the BH hamiltonian (section III of SI). First, we deduced that the 800 nm period lattice is mostly bound to the regime of vanishingly small tunnelling, whereas for the 400 nm period device excitons can efficiently explore lattice sites within our experimental timescale (Fig.S10). On the other hand, Fig.S9 shows that $U$ does not exceed $\Delta E$ for most WS of the 800 nm device, while we find $U\lesssim\Delta E$ from the 5$^\mathrm{th}$ to 8$^\mathrm{th}$ WS for the 400 nm period lattice. Studying both devices then allows us to probe the interplay between $U$, $t$, $\Delta E$, and the specific WS where a MI possibly buildups in the multi-orbital regime.}

\fd{In both lattice potentials, for most WS  ($U/t$) theoretically exceeds largely the critical ratio $\sim$20 \cite{Trivedi_2011} for the realisation of a Mott phase (Fig.S11). The latter is then signalled by a uniform and integer filling of lattice sites, with excitons all confined in the same WS.} Furthermore, in the MI regime fluctuations of the number of excitons per site shall be strongly reduced, manifesting that the compressibility is minimised \cite{Chin_2009}. To evidence Mott phases we have to extract the occupation of WS across the lattice. Experimentally, this is achieved by studying the photoluminescence (PL) spectrum radiated by dipolar excitons. Indeed, each WS leads to an individual optical emission line at its corresponding energy, and whose amplitude translates into the fraction of excitons occupying the considered state.

Figure 2.a presents the spatially and spectrally resolved PL for the 800 nm period lattice. For these experiments the laser excitation was set to $P=6$ nW with a spot size of 4 $\mu m$. In this situation we find that lattice sites are in average filled with around 4 excitons (section II of SI). \fd{Relying on the calculated energy positions of WS and by only adjusting the mean occupation fraction $\bar{p}$ for each WS (section II.C of SI),} Fig.2.b shows that we quantitatively reproduce the spectrum emitted all along the vertical direction of the device. For the 3 positions highlighted in Fig.2.b, Fig.2.c reports the amplitude of $\bar{p}$ for all 15 WS. 

The spatial variations of the PL spectrum shown in Fig.2 reveal that excitons realise a normal fluid, since several WS are significantly occupied. This first results from the strength of on-site interactions, since for an average filling of around 4 excitons per site, the interaction strength (3$\cdot U$) greatly exceeds \fd{$\Delta E$} (Fig.S9). The exciton population must then be arranged between several WS in each site. On the other hand, let us underline that the energy relaxation in the lattice also contributes to setting the occupation fractions of WS. Indeed, dipolar excitons are non-resonantly injected, with a high kinetic energy at the termination of the loading laser pulse. Then, they rapidly thermalise to the bath temperature \cite{Ivanov_2004,Beian_2017}, within few ns, without being necessarily confined by the lattice yet. For that, excitons need to relax between WS which is more tedious since \fd{$\Delta E$} exceeds the thermal energy by over 3-fold. Two or many-body exciton collisions, combined with exciton tunnelling efficient for weakly confined states (Fig.S10), thereby constitute key mechanisms.

To explore the phase diagram of the BH hamiltonian, we varied the filling of lattice sites by sweeping the average power $P$ of the loading laser pulse.  Figure 3 shows that we thus found two specific values, namely $P= 2$ nW (Fig.3.a) and  $P=3$ nW (Fig.3.b) for which the PL spectrum essentially reduces to a single sharp emission line, extending over 3 (Fig.3.a) and 4 (Fig.3.b) lattice sites vertically. Note that along the horizontal direction we average here 3 lattice rows given the optical magnification of our experiments. Modelling the PL spectra, we deduce that in these experiments over 40$\%$ of confined excitons occupy the 7$^\mathrm{th}$ WS, whereas the other 14 accessible states are all very weakly populated (middle and bottom rows of Fig.3.a-b). Furthermore, the occupation of the 7$^\mathrm{th}$ WS displays weak spatial variations for both $P=$ 2 and 3 nW (insets in the bottom panels of Fig.3.a and 3.b). 

The average filling of lattice sites can not exceed 2 to 3 excitons for the measurements shown in Fig. 3.b (section II.B of SI). Noting that in the middle row of Fig.3.a-b we observe a ratio of around 2 between the peak intensity of the spectra, and since each emission is uniform spatially, we deduce a filling $n_X=1$ and 2 excitons in every lattice site for Fig.3.a and Fig.3.b respectively. Furthermore, we measure an energy shift equal to 80$\pm$15 $\mu$eV between the maximum of the spectra. This splitting  quantifies the magnitude of $U$, which energetically separates the configurations where lattice sites are uniformly filled with one and two excitons. Remarkably, our experiments agree with the value of $U$ theoretically expected for the 7$^{th}$ WS (80 $\mu$eV, see Fig.S9). 
Let us then note that the magnitude of $U$ is further supported by independent experiments performed at variable temperature (Fig.S12). Finally, Fig.3.c shows that while varying the average filling of the lattice, we find that the occupation of a single WS is only dominant for $P=(2,3)$ nW. This shows that otherwise excitons realise a normal fluid, as illustrated in Fig.2.

To confirm that Fig.3.a-b evidence Mott-like phases, we computed the standard deviation of the WS occupation, $\sigma(p(n))$, since $\sigma(p(n))/\overline{p(n)}$ is proportional to $(\kappa k_BT)^{1/2}$ where $\kappa$ denotes the compressibility \cite{Chin_2009}. Figure 3.d then shows that $\sigma(p(7))$/$\overline{p(7)}$ is strongly decreased for $P=(2,3)$ nW only. This behaviour signals that $\kappa$ is minimized, as expected for a MI, well below any other filling for every WS (grey region in Fig.3.d).

\fd{The measurements summarised in Fig.3 provide the fingerprints expected for Mott-like phases, with $n_X$=(1,2) excitons per lattice site in the 7$^{th}$ WS. Comparing these two realisations we measured the amplitude of U, thereby confirming that  $U\lesssim\Delta E$ while verifying the good accuracy of our theoretical calculations. From these we expect that $t\sim 5\cdot10^{-8} \mu$eV for the 7$^{th}$ WS. Our observations are then bound to the regime of vanishing tunnelling, where a MI is nevertheless theoretically expected \cite{Salomon_2010}.}

\fd{To verify that Mott phases exhibit identical signatures when $t$ is orders of magnitude larger, we repeated the same experiments for the 400 nm period device. Indeed, for this latter $t\sim$0.06 $\mu$eV in the 5$^{th}$ WS (Fig.S10), so that the tunnelling characteristic time between two sites is 15 times shorter than our 50 ns long measurement time. Let us then note that studying shorter period lattices inevitably yields increased magnitudes for $U$ since WS are more tightly confined (Eq.(32) of SI). Accordingly, $U\simeq185\mu$eV $\sim\Delta E$ for the 5$^{th}$ WS of the 400 nm period lattice.}

Figure 4.a reveals that we again found a specific excitation, $P$= 8.5 nW for a 8 $\mu$m spot, where 43$\%$ of excitons occupy the 5$^{th}$ WS. In these experiments the average filling is about 1 exciton per site (section II.B of SI). Since the occupation of the 5$^{th}$ WS is uniform over around 100 sites (inset in the bottom panel of Fig.4.a), we deduce that lattice sites are in fact all occupied by $n_X$=1 exciton. Moreover, Fig.4.b underlines that we do not observe any other average filling where the occupation of the 5$^\mathrm{th}$ WS is dominant. This behaviour is consistent with $n_X$=1, since the magnitude of $U$ is about the energy gap around the 5$^\mathrm{th}$ WS (Fig.S9). Accordingly, populating this WS with more than one exciton is not energetically favourable. In the same way we deduce  that in Fig.4.a exciton tunnelling is inhibited. Otherwise, doubly filled lattice sites would lead to a wider distribution of significantly occupied WS, because $U$ is comparable to \fd{$\Delta E$}. \fd{We then note that in Fig.4.a the PL energy is constant across the lattice within our instrumental resolution (15 $\mu$eV). The latter provides therefore an upper bound for the variation of WS energies between lattice sites. Accordingly, at T=330 mK possible lattice inhomogeneities are not sufficient to inhibit exciton tunnelling \cite{Mott}  (section III.G of SI).} 

As for the 800 nm period lattice, we finally computed the local standard deviation of the PL intensity, over a region extending around 20 sites at the center of the emission. Fig.4.c shows that $\sigma(p(5))$/$\overline{p(5)}$ is strongly reduced for $P$= 8.5 nW, compared to all other WS at any other average filling of the lattice. Figure 4 then signals a MI, with $n_X$=1 exciton per lattice site  in the 5$^{th}$ WS.

\fd{Comparing theoretical values of $U$ and $\Delta E$ for the different WS (Fig.S11), our observations suggest that Mott phases are formed in the lowest energy WS where U is smaller than $\Delta E$, for both lattices.} This condition is necessary to allow 2 excitons to occupy the same WS in one site, while ensuring that the potential energy is minimised and $(U/t)$ maximum. Thus, the parameter space where MI emerge is the widest. However, it is less clear why this condition also applies to Mott phases with $n_X$=1. This possibly relates to the precise structure of the phase diagram, where a variety of ordered phases may coexist in the multi-orbital regime for on-site interactions comparable or larger than the separation between WS. \fd{Shorter period lattices provide a model environment to test the presence of such ordered phases, since the number of WS is then reduced}. Decreasing the lattice period would also allow us to extend the BH hamiltonian to interactions between nearest neighbouring lattice sites. \fd{These can reach an amplitude around 35 $\mu$eV for 125 nm period devices, while $t$ possibly varies between 2 and 7 $\mu$eV by adjusting the lattice depth. Many-body phases breaking the lattice symmetry are then theoretically accessible for $T\sim$10 mK \cite{Capogrosso_2010}}.

\section*{Acknowledgments}
We would like to thank M. Holzmann for his crucial support to calculate microscopically dipolar interactions in the lattice, M. Lewenstein, M. Polini and A. Reserbat-Plantey for a critical reading of our manuscript, together with S. Gasparetto for graphical works. Our research has been financially supported by the Labex Matisse and by IXTASE from the French Agency for Research (ANR-20-CE30-0032-01). The work at Princeton University was funded by the Gordon and Betty Moore Foundation through the EPiQS initiative Grant GBMF4420, and by the National Science Foundation MRSEC Grant DMR 1420541.

\section*{Data availability}
The data supporting the studies presented here are available for download upon request.

\newpage

\newpage

\newpage

\clearpage

\pagebreak

\onecolumngrid

\begin{figure}[!ht]
  \includegraphics[width=\linewidth]{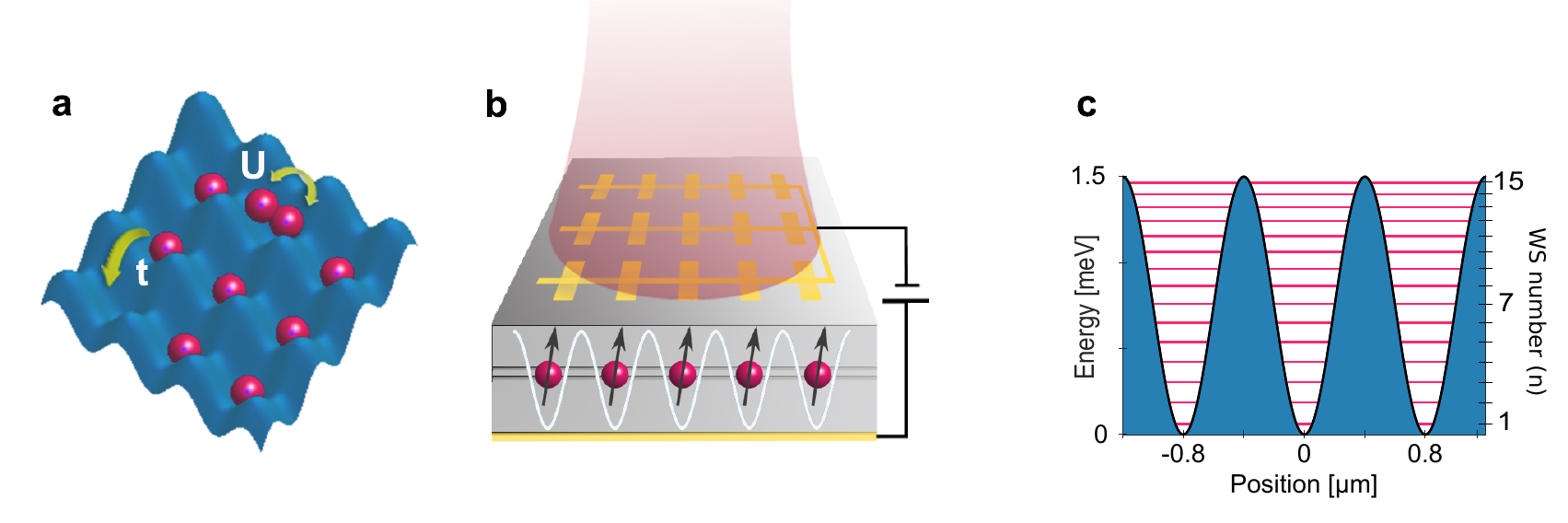}
  \caption{\textbf{Bose-Hubbard physics.} a) Within the BH model, bosonic particles (red) confined in a two-dimensional lattice (blue) experience on-site interactions with a magnitude $U$ while the tunneling strength between nearest neighbouring sites has an amplitude $t$. b) Our semiconductor device relies on two GaAs quantum wells, each confining electrons or holes that form dipolar excitons (red ball). These are confined in a two-dimensional electrostatic lattice due to the interaction between their permanent electric dipole (arrow) and the electric field imposed by the array of surface gate electrodes (gold). Electronic carriers are optically injected in the lattice using a laser beam focussed at the surface (red). c)  For a 1.5 meV deep lattice with \SI{800}{\nano\metre} period 15 Wannier states (red) coexist, separated by around 100 $\mu$eV.}
  \label{fig:fig1}
\end{figure}

\begin{figure}[!ht]
  \includegraphics[width=.9\linewidth]{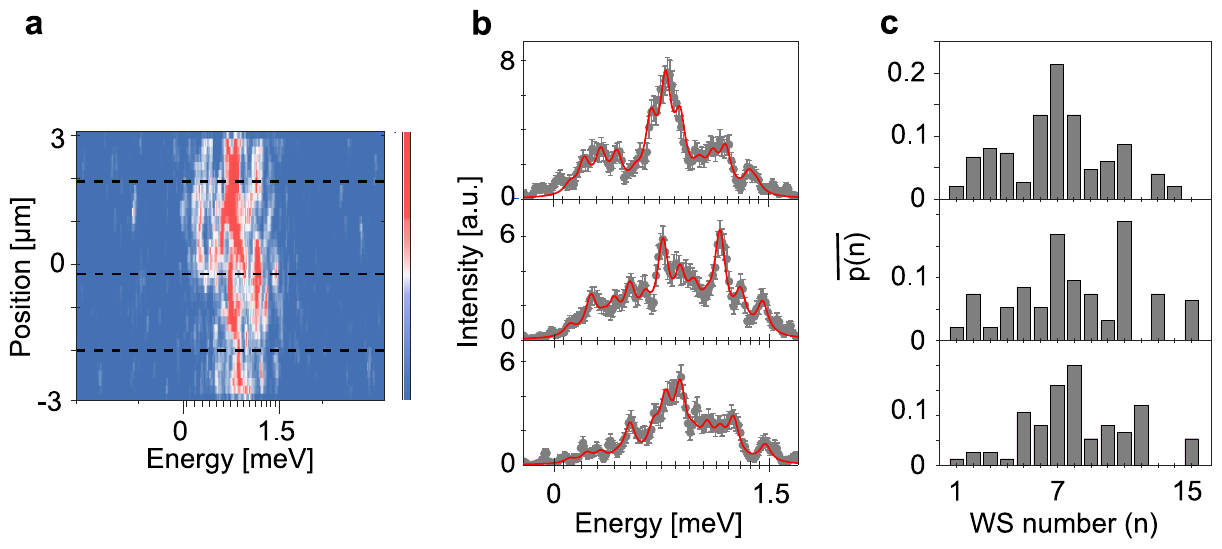}
  \caption{\textbf{Arbitrary filling of a 800 nm period lattice.} a) Spatially and spectrally resolved photoluminescence when we impose an average filling around 4 excitons per lattice site ($P=6$ nW). b) Average spectra measured at the positions underlined by the dashed lines in the panel a), from top to bottom. Each spectrum is obtained by averaging  over a 1.5 $\mu$m region vertically which corresponds to our optical resolution. Experimental data are displayed by gray points, error bars representing the poissonian noise, while the solid red lines show the modelled photoluminescence spectra. c) Mean occupation fraction $\overline{p}$ of all 15 WS used to reproduce the experiments shown in b).}
  \label{fig:fig_2}
\end{figure}

\begin{figure}[!ht]
\centerline{\includegraphics[width=\linewidth]{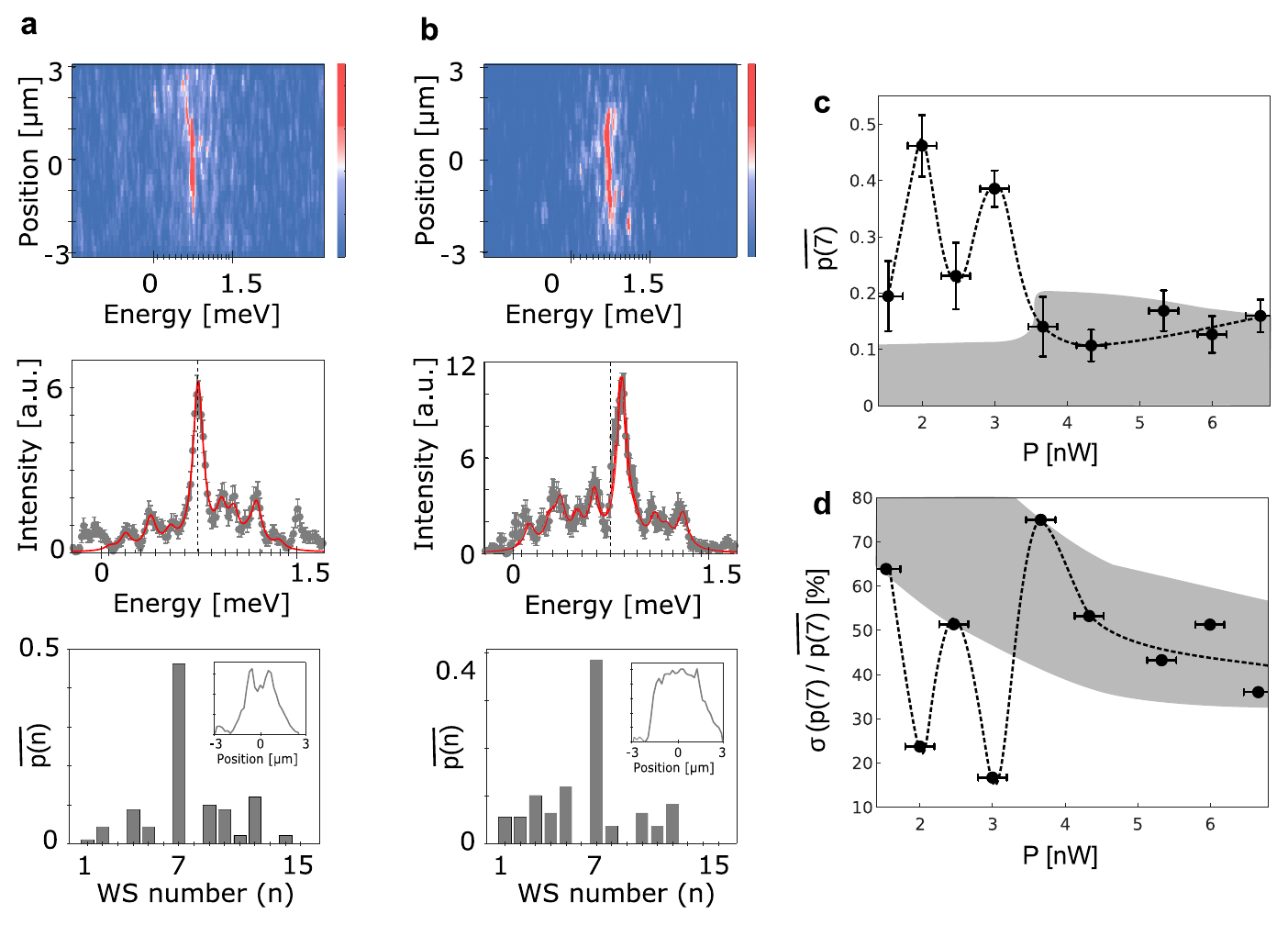}}
  \caption{\textbf{Mott-like phases in a 800 nm period lattice.} a) Spatially and spectrally resolved photoluminescence for $P=2$ nW  (top).  The middle panel displays the spectrum in the 1.5 $\mu$m central region (gray points) together with the fitted profile (red) from which the mean occupation fraction $\overline{p}$ of all WS is deduced (bottom panel, where the inset shows the spatial variation of  $\overline{p(7)}$). b) Same measurements as in a) but for $P=3$ nW. The vertical dashed line in the middle panel marks the maximum of the spectrum for $P=2$ nW. In a) and b) the horizontal ticks indicate the energies of WS in the top and middle panels. c) Mean fraction $\overline{p(7)}$ as a function of $P$ where \fd{vertical} error bars display the standard deviation  \fd{while horizontal error bars provide the stability of the laser power}. The gray region delimits the occupations of all other WS. d) Normalised standard deviation for the occupation of the 7$^\mathrm{th}$ WS as a function of $P$ (dark points). The gray shadow area delimits the region where the normalised standard deviation of all the other WS is bound.}
    \label{fig:fig_3}
\end{figure}

\newpage

\begin{figure}[!ht]
  \includegraphics[width=.75\linewidth]{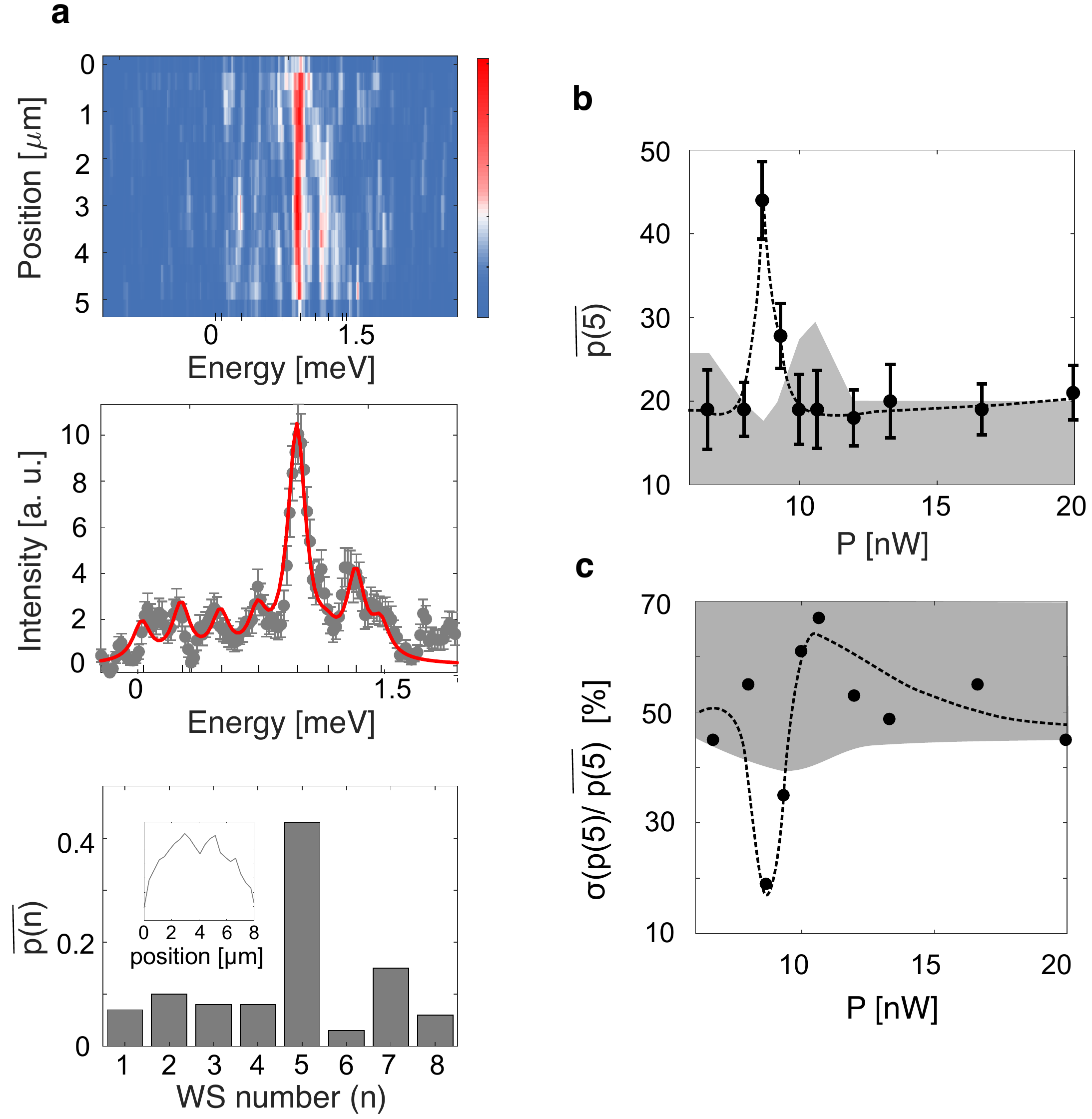}
  \caption{\textbf{Mott insulator in a 400 nm period lattice.} a) Spatially and spectrally resolved photoluminescence for a 8 $\mu$m wide excitation spot and $P=8.5$ nW (top).  The middle panel displays the average spectrum (gray points) together with the fitted profile (red) from which the mean occupation fraction $\overline{p}$ of all WS is deduced (bottom panel, where the inset shows the spatial variation of $\overline{p(5)}$). In the top and middle panels the horizontal ticks indicate the energies of WS. b) Mean occupation fraction $\overline{p(5)}$ as a function of $P$. \fd{Vertical} error bars display the standard deviation  \fd{while horizontal error bars overlap with the dot size}. The gray region delimits the occupation fraction of all other WS. c) Normalised standard deviation of $\overline{p(5)}$ as a function of $P$, in the 1.5 $\mu$m region at the center of the image a) (dark points). The gray shadow area delimits the region where the normalised standard deviation of all the other WS is bound.}
\end{figure}

\clearpage

\newpage
\pagebreak

\newpage
\newpage
\pagebreak

\clearpage

\centerline{\textbf{\large{Supplementary Informations}}}

\section{Electrostatic lattices}

Our two devices rely on a pair of 8 nm wide GaAs quantum wells (DQW), separated by a 4 nm Al$_{.3}$Ga$_{.7}$As barrier, and embedded in Al$_{.3}$Ga$_{.7}$As. The centers of the two quantum wells are thus separated by $d$= 12 nm which sets the magnitude of the permanent electric dipole carried by excitons made by one electron confined in one quantum well and a hole confined in the other quantum well, i.e. dipolar excitons. To engineer a periodically varying field in the plane of the DQW we control the geometry of metallic electrodes deposited at the surface of our devices, as well as the positions of the DQW in the heterostructure. Thus, we aim at ensuring that the magnitude of the electric field component perpendicular to the DQW, $E_z$, is sufficiently large to confine excitons. Indeed, excitons have an electric dipole oriented perpendicular to the DQW plane so that their potential energy scales as ($-edE_z$), $e$ denoting the electron charge. Also, we have to ensure that the electric field component in the plane of the DQW, $E_{x,y}$, is minimised to prevent exciton dissociation.

\subsection{Geometry of sample structures}

As shown in Table 1 and Fig.S1, we systematically placed the DQW at h=150 nm above a conductive $n$-doped GaAs layer that serves as electrical ground. On the other hand, to engineer 800 nm and 400 nm period lattices we varied the distance H between the DQW and the surface, H=450 nm and H=200 nm respectively. Thus, H is kept small compared to the lattice period that defines the characteristic size of the electrode pattern imprinting the electric field perpendicular to the DQW. As further discussed in the following, the profile of $E_z$ is then efficiently varied in the plane of the DQW, and so is then the excitons confinement.

\begin{center}
   \begin{tabular}{  | l | c | c | }
     \hline
		            \multicolumn{3}{|c|}{ Device design} \\
  \hline
          Lattice period (a) & \SI{800}{\nano\metre}  & \SI{400}{\nano\metre}  \\ \hline
DQW to surface (H)& \SI{450}{\nano\metre}& \SI{200}{\nano\metre} \\ \hline
DQW to ground (h) & \SI{150}{\nano\metre}& \SI{150}{\nano\metre} \\ \hline
Distance from centers of the two QWs (d)&\multicolumn{2}{c|}{ \SI{12}{\nano\metre}}\\  \hline
Rectangle width ($r_W$)& \SI{320}{\nano\metre} & \SI{170}{\nano\metre}\\     \hline
Rectangle height ($r_H$)&\SI{580}{\nano\metre} &\SI{330}{\nano\metre}\\     \hline
Wire thickness ($w_H$)&\SI{100}{\nano\metre} & \SI{90}{\nano\metre} \\     \hline
Separation distance ($s_H$)&\SI{220}{\nano\metre} & \SI{70}{\nano\metre} \\     \hline
\end{tabular}
\end{center}
\textit{\textbf{Table 1}: Geometric parameters of the devices implementing 800 nm and 400 nm period lattices. These parameters are all illustrated in Fig.S1.}

\centerline{\includegraphics[width=.85\linewidth]{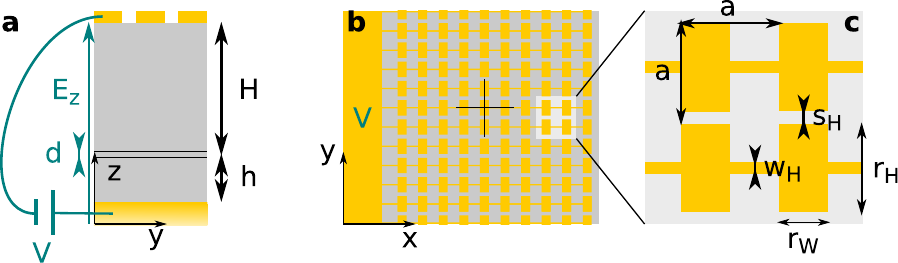}}
\textit{\textbf{Fig. S1}: a) Sketch of the section of our field effect devices where a DQW is positioned at a distance h from a grounded surface and H from the surface where metallic electrodes are deposited and polarised at a potential V. Thus we impose an electric field with a magnitude $E_z$ perpendicular to the DQW. b-c) Geometry of the electrode arrangement deposited at the surface of our devices.}
\vspace{.5cm}

We performed finite element simulations in order to design the appropriate electrode patterns realising 400 and 800 nm period electrostatic lattices for dipolar excitons. We first concluded that our device needed to rely on a single interdigitated electrode, which once polarised imprints the lattice sites whereas potential barriers are induced by the unpolarised interstices between the electrodes. This configuration differs from one of our previous studies \cite{Lagoin_2020,Lagoin_2021} where 2 electrodes were used, one defining the lattice sites and one the barriers. In fact, for lattice periods below 1 $\mu$m the geometrical factors hardly allow one to use two interdigitated electrodes. As illustrated in Fig.S1.b-c we precisely relied on an arrangement of rectangles to define the lattice sites, with dimensions (r$_H$, r$_W$), electrically connected by a wire with a width w$_H$. Finally, note that each row of electrodes is separated by a distance s$_H$ and connected to a large and continuous electrode brought to a potential V. 

Table 1 displays the values of all geometrical factors that we used for the 400 nm and 800 nm period lattice potentials. The corresponding electrode patterns were then imprinted at the surface of our device using electron-beam lithography followed by metal deposition. As illustrated in Fig.S2 electron microscope images demonstrated that the designs were successfully realised with a precision of around 10 nm.

\vspace{.5cm}
\centerline{\includegraphics[width=.7\linewidth]{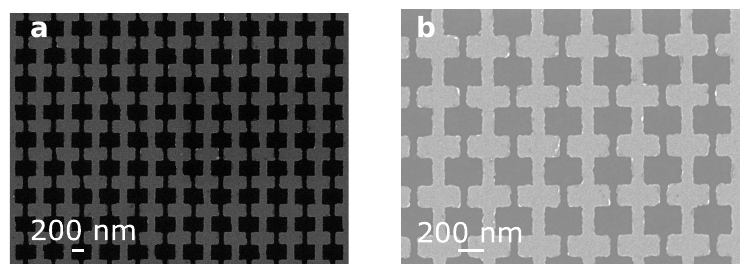}}
\textit{\textbf{Fig. S2}: Electron microscope image of surface electrodes defining the 800 nm a) and 400 nm b) period lattice potential.}
\vspace{.5cm}

\subsection{Lattice depth and internal electric fields}

The electric fields in and out of the DQW plane, $E_{x,y}$ and $E_z$ respectively, have amplitudes directly depending on the potential V applied to the surface electrodes. Experimentally, we measured the magnitude of $E_z\sim$$V/(H+h)$ by comparing the energy of the photoluminescence emission of dipolar excitons to the one of direct excitons confined in each of the quantum wells. Indeed, the energy difference between these two emissions is given by $e.d.V/(H+h)$. Setting $V$=-2 V and $V$=-1.4 V for the 800 nm and 400 nm lattice devices, we observed  that dipolar excitons radiate a photoluminescence at 1.527 and 1.520 eV respectively, whereas the direct excitons photoluminescence lied at 1.577 eV. We thus confirmed that the amplitude of the internal electric field $E_z$ is  exactly given by the potential applied on the surface electrodes, for both structures.

In Fig.S3 we present the spatial profiles of $E_z$ and $E_{x,y}$ computed for the two devices studied in the main text, for V= -2V and -1.4 V respectively. Fig.S3.a-b show that $E_z$ varies sinusoidally in the plane of the DQW. It is centred at 3.08 and 3.86 V/$\mu$m for the 800 nm and 400 nm devices respectively, with a modulation amplitude around 0.1 V/$\mu$m. On the other hand, $E_{x,y}$ is centred at 0 V/$\mu$m and also varies periodically along the two axis of the lattices, with an amplitude 0.1 V/$\mu$m. These simulations first verify that $E_{x,y}$ has an amplitude that does not exceed 3$\%$ of the one of $E_z$ for both devices. Accordingly, the in-plane electrostatic force induced on dipolar excitons does not exceed 1 meV whereas the excitons binding energy is about 3-4 meV. This ensures that excitons are not ionised by in-plane fields in our studies. Moreover, let us note that $E_{x,y}$ is vanishing in the region at the centre of the lattice sites, i.e. where $E_z$ is the largest. In fact, in-plane fields are largest at the positions of the electrostatic barriers, which further reduces the role of exciton ionization in our experiments.

The profile of the lattice depth where dipolar excitons are confined is presented in Fig.S3.c-d for the 2 devices. We observe that the two lattice potentials have a confinement depth of about 1.5 meV. To impose these lattice depths note that surface electrodes are polarised at V= -2V and -1.4V for the 800 nm and 400 period lattices respectively. This is necessary given that $H$ is approximately doubled for the former structure compared to the latter one.

\vspace{.5cm}
\centerline{\includegraphics[width=.8\linewidth]{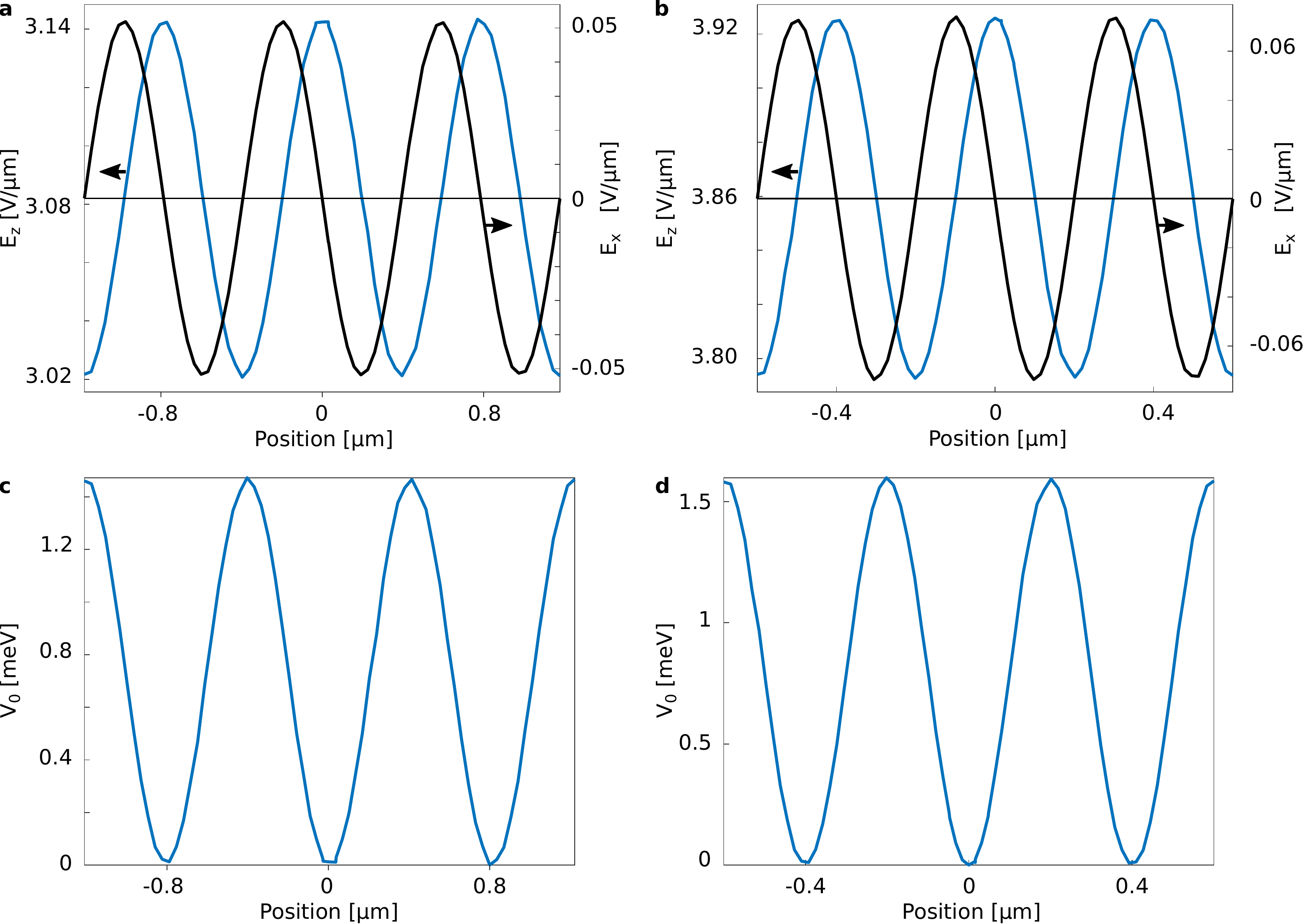}}
\textit{\textbf{Fig. S3}: a-b) Spatial profile of out and in-plane electric fields, $E_z$ (blue) and $E_{x,y}$ (black) respectively, simulated for the 800 nm (a) and 400 nm period (b) lattices. c-d) Spatial profile of the lattice depth $V_0$ deduced from the amplitude of $E_z$ for the 800 nm (c) and 400 nm period (d) lattices.}
\vspace{.5cm}

\section{Experimental procedure}

\subsection{Stroboscopic spectroscopy}

Figure 1.b illustrates that excitons are optically injected in the lattice within a gaussian laser excitation spot with a full width at half maximum of 4 $\mu m$ for the 800 nm lattice and 8 $\mu m$ for the 400 nm lattice. Figure S4.a illustrates that this is realised non-resonantly, by exciting the direct excitons (DX) transition of each quantum wells (at $\sim$787 nm). Electrons and holes are then directly injected in the two quantum wells and not throughout the field-effect devices. Dipolar excitons (IX) are formed once electrons and holes have tunnelled towards their minimum energy states, lying in a distinct layer since our heterostructure is electrically polarised. For this laser excitation, the photo-induced current is then minimised and bound to 10-20 pA, so that  the density of photo-induced free carriers is also minimised. Moreover, note that for such non-resonant excitation dipolar excitons are initially characterised by a high kinetic energy and thermalise to the bath temperature in typically less than 10 ns. At the termination of the laser excitation, excitons are then not confined by the lattice potential.

Fig.S4.b illustrates that our studies rely on a pulsed laser excitation with 100 ns duration, repeated at a frequency equal to 850 kHz. The photoluminescence emitted by dipolar excitons is then analysed at variable delays to the laser pulse, in a time window 50 ns long. In the measurements discussed in the main text the delay is fixed to 250 ns while the average power of the loading laser pulse $P$ is varied. In this situation the photoluminescence radiated by direct excitons is vanishing since these are characterised by an optical lifetime in the ps time domain. 

Importantly, the experiments presented throughout our manuscript are performed stroboscopically. Precisely, for every experimental settings we perform 10 successive measurements under fixed conditions. A single stroboscopic measurement in practice requires about a 30 seconds long acquisition, so that the spatially and spectrally resolved photoluminescence images, as shown in Fig.2-4, require a 5 minutes long measurement. These images correspond then to a statistical average over 255$\cdot$10$^{6}$ realisations.  

For the measurements shown in Fig.2-4 photoluminescence spectra are acquired with a 1800 lines/mm grating, so that the emission energy is sampled with a precision equal to \SI{15}{\micro\eV}. Also, we observe that the response function of our spectrometer is well adjusted by a lorentzian line with \SI{100}{\micro\eV} line-width, as possibly measured with e.g. a Mercury (Hg) emission line. When fitting the photoluminescence spectra shown in Fig.2-4 we then assigned this 100 $\mu$eV lorentzian profile to the emission of each WS.

\vspace{.5cm}
\centerline{\includegraphics[width=.9\linewidth]{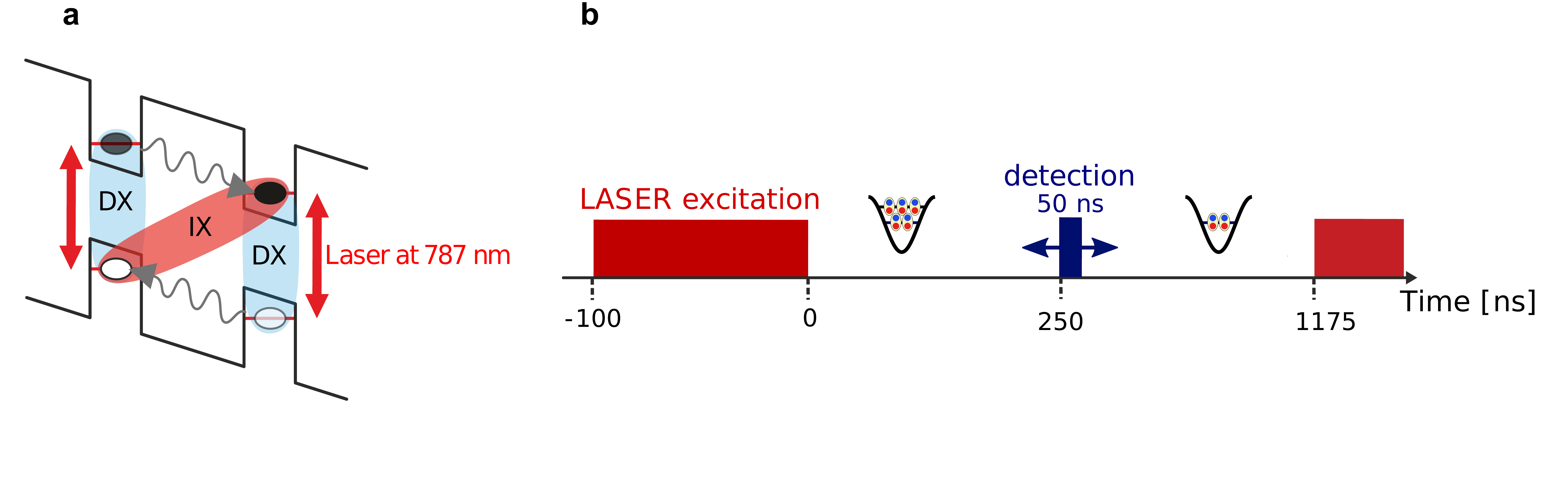}}
\textit{\textbf{Fig. S4}: a) Schematic energy levels of an electrically polarised DQW. Direct excitons (DX) are made of electrons and holes confined in the same quantum well while dipolar excitons (IX) are made of spatially separated electrons and holes, since their minimum energy states lie in a distinct layer.  b) Our measurements are all performed dynamically, relying on a 100 ns long laser excitation repeated at 850 kHz, while the excitons' photoluminescence is detected in a 50 ns long time window at a variable delay after extinction of the laser excitation. In the experiments discussed in the main text the delay is always set to 250 ns while the average power of the loading laser pulse $P$ is varied.}
\vspace{.5cm}

\subsection{Estimation of the exciton density}

To evaluate the average exciton density in the lattice we measured the decay of the photoluminescence energy after extinction of the loading laser pulse. Precisely, we extracted the photoluminescence blueshift $\Delta E_X$, i.e. the difference between the photoluminescence energy 250 ns after extinction of the laser pulse, to the one for much longer delays ($>$ 500 ns) when the density is vanishingly small \cite{Beian_2017,Lagoin_2020}, see Fig.S6.a. Indeed, in our experiments the photoluminescence energy $E_X$ is given by the sum of the excitons potential energy ($-e.d.E_z$) and the repulsive dipolar interaction between excitons. The latter approximately scales as $u_0 n_X$ where $n_X$ denotes the average exciton density for a homogeneous fluid. The parameter $u_0$ is most accurately computed by taking into account a depletion region around each exciton representing the very strong dipolar interactions between excitons at short distances \cite{Ivanov_2010,Schindler_2008,Rapaport_2009}. Figure S5 presents the theoretical scaling of the photoluminescence blueshift as a function of the average exciton density.     

\vspace{.5cm}
\centerline{\includegraphics[width=.4\linewidth]{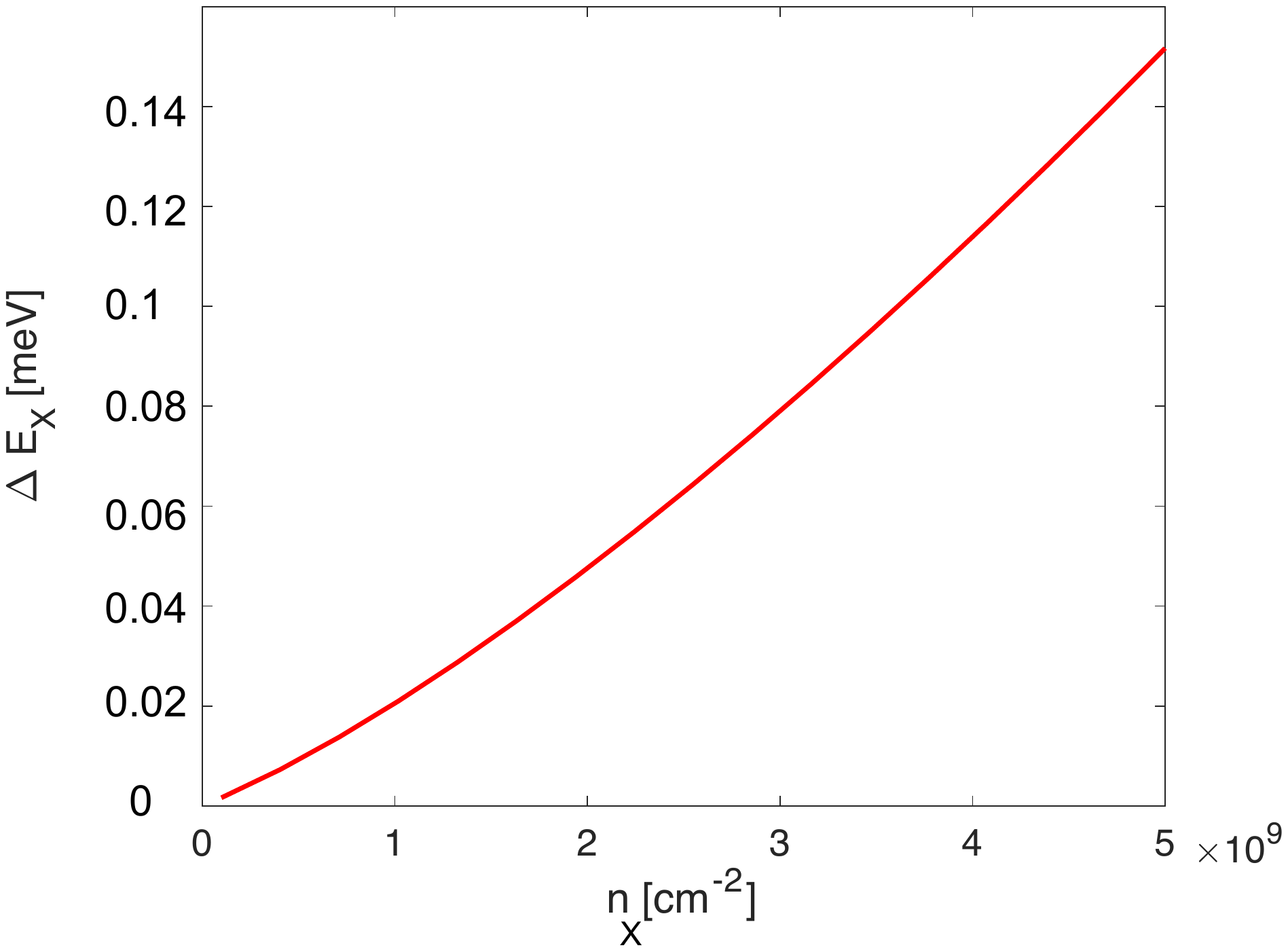}}
\textit{\textbf{Fig. S5}: Blueshift of the photoluminescence energy as a function of the exciton density $n_X$ calculated for a homogeneous spatial distribution.}
\vspace{.5cm}

To calibrate the average density of excitons injected by our laser excitation we used a region of our devices where $E_z$ is uniform, i.e. under a non patterned surface electrode, so that the resulting exciton density is homogeneous spatially. Indeed, the variation of the photoluminescence blueshift $\Delta E_X\propto u_0\cdot n_X$ (Fig.S5) only applies to a spatially homogeneous fluid. Furthermore, to deduce in the most accurate way $n_X$ we used a mean laser power $P$ larger than the one for which Mott phases emerge. Thus, $n_X$ is larger and so is then $\Delta E_X$. Figure S6.a provides a concrete example for the heterostructure used to engineer the 800 nm period lattice and for $P=6.5$ nW. There we note that $\Delta E_X$ does not exceed 100 $\mu$eV for a delay set to 250 ns.  Accordingly, we deduce that the average exciton density is around 3 10$^9$ cm$^{-2}$ in these measurements. Finally, Fig. 6.b presents the decay of the integrated intensity of the photoluminescence for these experiments. It shows that dipolar excitons exhibit a radiative lifetime of around 700 ns. 

\vspace{.5cm}
\centerline{\includegraphics[width=.7\linewidth]{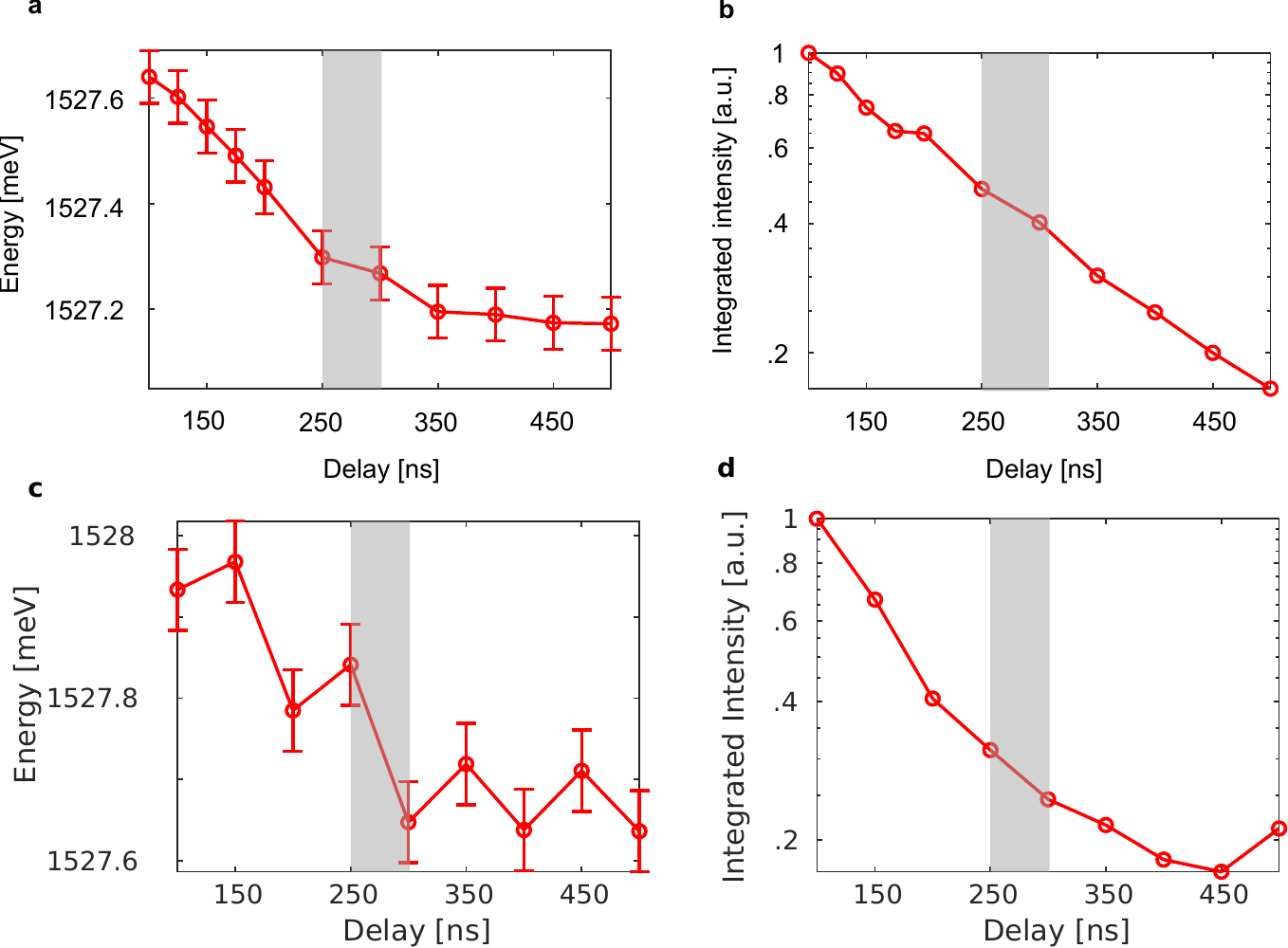}}
\textit{\textbf{Fig. S6}:  a) Energy and b) integrated intensity of the photoluminescence measured as a function of the delay to the end of the loading laser pulse, for a homogeneous fluid of dipolar excitons. The laser excitation average power is set to $P\sim6.5$ nW which corresponds to the highest excitation in Fig.3.c. For a delay set to 250 ns (gray region), as in the experiments shown in Fig.2-4, the panel a) shows that the blueshift of the photoluminescence energy does not exceed 100 $\mu$eV. c-d) Same measurements performed for the heterostructure where we realise the 400 nm period lattice. The laser excitation has this time a full-width-at-half-maximum around 8 $\mu$m, and in these experiments the average power is set to $P\sim17$ nW, i.e. twice that of Fig.4.a.}

\vspace{0.5cm}

For the experiments reported in Fig.2-3 for which the excitation spot was gaussian with around 4 $\mu$m full-width-at-half-maximum at the surface of the device, Fig.S6.a shows that the average exciton density is around  3 10$^9$ cm$^{-2}$ for $P$= 6.5 nW. In the lattice potential, at this density the number of excitons per lattice site is around 4 for the 800 nm period device. Since we verified experimentally that the integrated intensity of the photoluminescence scales linearly with the laser excitation power, we conclude that for $P$= 3 nW the average filling of the lattice potential is about 2 excitons per site. For the 400 nm period device studied in Fig. 4 we performed the same calibration procedure, as illustrated in Fig.S6.c-d. We concluded that for a delay set to 250 ns, for $P$=17 nW, this time for an excitation laser spot with around 8 $\mu$m full-width-at-half-maximum, the average density is about 3-4 10$^9$ cm$^{-2}$ corresponding to 2 excitons per lattice sites in average. For P=8.5 nW, as in Fig.4.a, we then expect that lattice sites are in average filled with one exciton. 

To conclude, we would like to underline that one must rely on the photoluminescence blueshift to extract the average density of dipolar excitons. Performing this measurement by monitoring the dynamics of $E_X$ after a laser excitation constitutes to the best of our knowledge the most reliable approach. Alternatively one may vary the power of the laser excitation to increase the density and then induce a blueshift of the photoluminescence, but in this case the strength of $E_z$ may vary together with the exciton density, notably due to photo-injected carriers trapped at the hetero-interfaces of the field-effect device \cite{Alloing_PhD}. Such variation of $E_z$ naturally blurs the deduced $n_X$, since both the excitons potential energy and the density induced blueshift vary. Therefore we discarded this second approach. 

\fd{\subsection{Reproduction of photoluminescence spectra}}

\fd{As previously mentioned, to reproduce the photoluminescence spectra displayed in Fig.2-4 we assign a lorentzian emission profile for each WS, with 100 $\mu$eV full-width-at-half-maximum (FWHM). Indeed this profile efficiently reproduces the spectral response of our imaging spectrometer. The photoluminescence spectra radiated by our lattice devices are then reproduced by summing 100 $\mu$eV FWHM lorentzian lines, one for each WS, with an amplitude controlled by the mean occupation fraction $\bar{p}$ of the considered state. $\bar{p}$ then constitutes the only adjustable parameter to reproduce the spectra displayed in Fig.2-4.}

\fd{Fig.S7 presents the residuals of our analysis (red), i.e. the difference between our measurements and the computed profile, for each spectrum shown in Fig.2-4. Thus, we verify that residuals are mostly bound to the statistical noise of our measurements (grey area). In fact, modelled spectra most significantly deviate from measured values when the photoluminescence signal is weak, i.e. when the signal-to-noise ratio is minimal. This situation occurs when WS are weakly populated.}

\vspace{.5cm}
\centerline{\includegraphics[width=.8\linewidth]{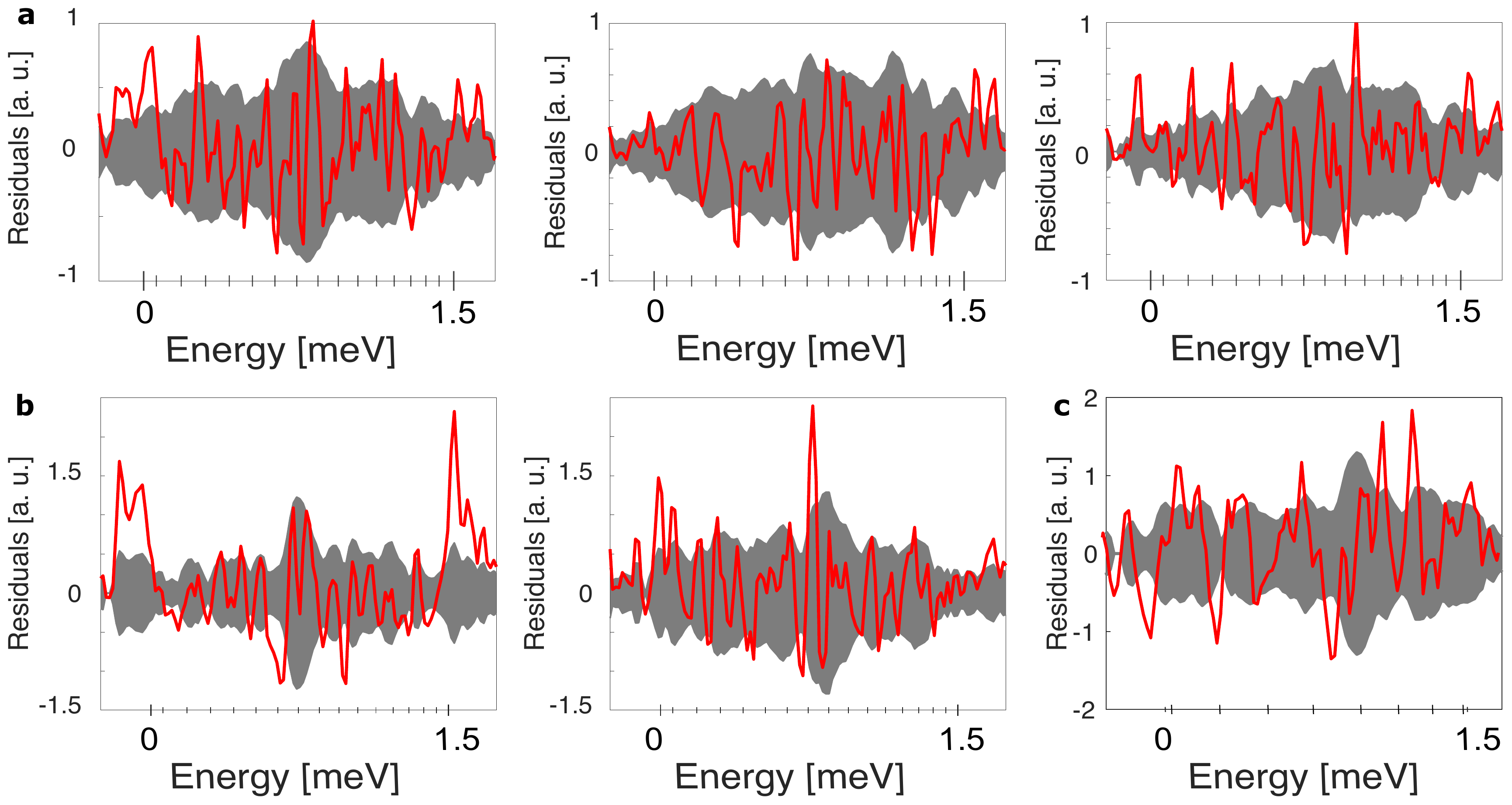}}
\textit{\fd{\textbf{Fig. S7}:  Difference between measured and calculated photoluminescence spectra (red) for the experiments shown in Fig.2 (a), Fig.3.a-b (b) and Fig.4.a (c). For all panels the grey area marks the statistical noise of our measurements. Horizontal ticks mark the positions of WS energies.}}

\vspace{.5cm}

\section{Bose Hubbard model for dipolar excitons}
\subsection{Two dimensional lattice potential}
\fd{Let us consider non interacting bosons in a two dimensional square lattice with an amplitude $V_0$ and a  sinusoidal profile in the  $(x,y)$ plane
\begin{equation}
V(x,y)=V(x)+V(y)=V_0 \sin^2{(Qx)}+V_0 \sin^2{(Qy)}
\end{equation}
with $Q=\pi/a$, where $a$ is the lattice period.
\\
In this situation, the $x$ and $y$ components of the hamiltonian are decoupled, so that the eigenvectors factorise as $\phi^{n}_q(x)\cdot\phi^{n}_\kappa(y)$. Restricting then the analysis to the $x$ direction only, the Schroedinger equation reads}



\fd{\begin{equation}
h(x,p_x)\phi^{n}_q(x)=E^{n}_q\phi^{n}_q(x)
\end{equation}
where $h$ denotes the non-interacting hamiltonian while the eigenvectors $\phi^{n}_q(x)$ are Bloch wave functions, i.e. product of a plane wave $\exp{(iqx)}$ with wavevector $q$, and a function $u^{n}_q(x)$ with the same periodicity as the periodic potential $V(x)$. 
$E^{n}_q$ are the eigenenergies, i.e. the energies of the bands with index $n$ associated with the potential $V(x)$ (see e.g. Ref.\cite{Greiner_PhD} for details).}
\\

\fd{Figure S8 a) and d) show the Bloch band energies for the $a$=800 nm and $a$=400 nm period lattices, as a function of the quasi wavevector $q$ in ($\pi/a$) unit of the first Brilloin zone. Figure S8 shows that 15 and 8 bands have an energy below the 1.5 meV barrier height of the 800 nm and 400 nm lattice potentials respectively. These are then confined inside the lattice potentials. We also observe that Bloch bands are energetically separated by around 100 and 200 $\mu$eV for the two devices respectively, most of them exhibiting a rather flat dispersion manifesting a deep confining potential.}
\\

\fd{\subsection{Wannier functions}
The characteristic kinetic energy of an exciton with an effective mass $m_x$ confined in the lattice potential is called the recoil energy $E_r=\frac{\hbar^2Q^2}{2m_x}$. In the lattice potential excitons remain mostly localized around one of the minima of the potential if the confining potential depth is much larger than the recoil energy i.e. $s=V_0/E_r\gg1$. 
The two devices that we study in the main text fulfill this condition (s=565 and 142 for the 800 and 400 nm period lattices respectively). Instead of Bloch wave-functions, we then consider Wannier functions that are exponentially localised around one potential minimum, i.e. one lattice site $x_j=j a$. Wannier functions are defined by
\begin{equation}
w^{n}_j(x)=\sqrt{\frac{a}{2\pi}}\int_{-\pi/a}^{\pi/a}\phi^{n}_q(x)\exp{(-ix_jq)}dq
\end{equation}
Fig.S8 c) and f) show the spatial profiles of Wannier functions of particular interest for the two lattices. As expected, we note that the degree of localisation decreases when the energy of Wannier states increases.}\\

\fd{\subsection{Non interacting hamiltonian}
%
In second quantisation, the non interacting Hamiltonian reads
\begin{align}
h&=\sum_n \sum_{j,l} J_n(j-l) \hat{a}^{n\dagger}_l\hat{a}^{n}_j \\
J_n(j-l)&=\frac{a}{2\pi}\int_{-\pi/a}^{\pi/a}E^{n}_q  \exp{(i(j-l)aq)} dq
\label{tunnel}
\end{align}
where $\hat{a}^{n}_j$ annihilates a particle in the Wannier function $w^{n}_j(x)$.}

\fd{Eq.(4) describes the hopping from a site $j$ to a site $l$ with a tunnelling amplitude $J_n(j-l)$, 
which depends on the band $n$ and on the distance $|j-l|a$ between the two sites. $J_n(j-l)$ is in fact equal to the matrix element of the hamiltonian between two Wannier functions
\begin{equation}
 J_n(j-l)=\int w^{n*}_j(x)(-\frac{\hbar^2}{2m_x}\frac{d^2}{dx^2}+V(x))w^{n}_l(x)dx
\end{equation}
As Wannier functions decay exponentially when $|x-x_j|$ is large, their overlap drops rapidly when $|l-j|$ increases, so that we only take into account nearest neighbour hopping $J_n(1)=-t_n$, which is symbolised by the sum $<j,l>$ in Eq.(7). We finally extend the one dimensional result to two dimensions. For our square lattice, the number of nearest neighbour is $z=4$ defined as lattice connectivity, and the hamiltonian reads
\begin{equation}
H=\sum_n \sum_{<j,l>} -t_n \hat{a}^{n\dagger}_l\hat{a}^{n}_j 
\end{equation}}

\subsection{Wannier energies}
$J_n(0)$ provides the energy associated to a given Wannier function localised on a site $j$, $w^{n}_j(x) \equiv  w^{n}_0(x)$.
For a given site $j$ there are $n$ Wannier states with energies $J_n(0)$ that correspond to the Bloch bands energies averaged over the first Brillouin zone
\begin{equation}
J_n(0)=\frac{a}{2\pi}\int_{-\pi/a}^{\pi/a}E^{n}_q dq
\end{equation}
Energies of Wannier states are represented in Fig.S8.b and S8.e for the 800 nm period and 400 nm period lattices respectively.

\centerline{\includegraphics[width=\linewidth]{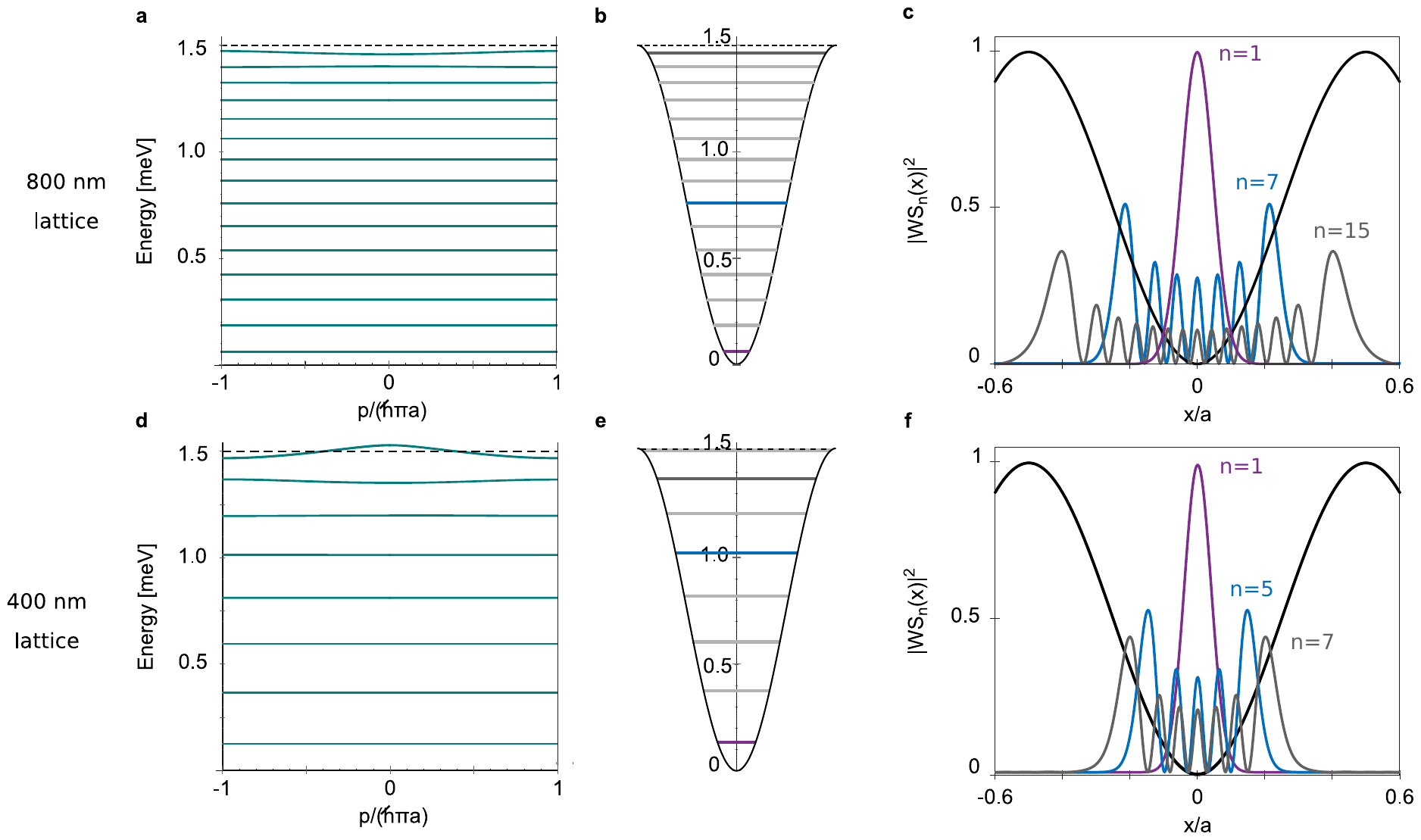}}
\textit{\textbf{Fig. S8}:  Dispersion of Bloch bands in the first Brillouin zone of the 800 nm period (a) and 400 nm period (d) lattice potential. The quasi-momentum $p=\hbar q$ is shown in units of $\hbar\pi/a$, where $a$ is the period of the lattice.  b), e) Energy of the Wannier states deduced from the Bloch bands. c) Spatial profile of the Wannier functions for the 1$^\mathrm{st}$ (violet), the 7$^\mathrm{th}$ (blue) and the 15$^\mathrm{th}$ (gray) confined states for the  800 nm lattice. f) Spatial profile of the Wannier functions for the 1$^\mathrm{st}$ (violet), the 5$^\mathrm{th}$ (blue) and the 7$^\mathrm{th}$ (gray) confined states for the 400 nm lattice. The horizontal axis marks the coordinate along one direction of the lattice, normalised by the period $a$.}\vspace{.25cm}

\subsection{Interaction term of the Bose-Hubbard hamiltonian}

The on-site two body interaction in a two dimensional lattice reads
\begin{equation}
U_{n}=\int \mathbf{dr_1dr_2} U_n(\mathbf{r_1}-\mathbf{r_2}) |w^n_0(\mathbf{r_1})|^2|w^n_0(\mathbf{r_2})|^2
\end{equation}
For excitons of DQW, the dipolar interaction between nearest neighbouring lattice sites is possibly approximated by  $V_1=\frac{(e\cdot d)^2}{4 \pi \epsilon a^3}$ \cite{Dutta_2015}, where $\epsilon=12.5 \cdot \epsilon_0$ is the Gallium Arsenide permittivity, $\epsilon_0$ being the vacuum permittivity. \textbf{$V_1$} is small for our lattice geometries compared to the on-site dipolar interaction (see Table 2) so that we neglect it in the following. This leads to a two-body on-site interaction term in the hamiltonian, namely 
\begin{equation}
H_{int}=\sum_{n,j}\frac{U_n}{2} \hat{n}_{j,n}(\hat{n}_{j,n}-1)
\end{equation}
where $\hat{n}_{j,n}$ is the particle number operator in the WS $n$, localised on a site $j$. In the following, we only consider interactions between particles in the same Wannier state for simplicity. Nevertheless, let us note that on-site interactions between particles occupying distinct Wannier states are in principle also to be taken into account. Let us also underline that Eq.(10) is possibly reduced to the fundamental state (n=1) only, if the population of higher WS is vanishing. This implies that the interaction strength $U$ is small compared to the energy separation between WS, which is not verified for our experiments.
\\
Taking into account nearest neighbour hoping and two body on-site interactions we finally obtain the Bose Hubbard Hamiltonian  
\begin{equation}
H_{BH}=H+H_{int}=\sum_n \sum_{<j,l>} -t_n \hat{a}^{n\dagger}_l\hat{a}^{n}_j +\sum_{n,j} \frac{U_n}{2}\hat{n}_{j,n}(\hat{n}_{j,n}-1)-\mu_0 \sum_{n,j} \hat{n}_{j,n}
\end{equation}
where the term $-\mu_0 \sum_{n,j} \hat{n}_{j,n}$ has been added to set the mean particle number, $\mu_0$ being the chemical potential.
\\

\subsection{On-site dipolar interactions}

\fd{To evaluate on-site interactions between excitons we neglect the role played by the excitons composite nature \cite{report}. Then, we assume that the mean spatial separation between excitons amounts to a few Bohr radii. Accordingly, we consider the dipolar potential} $U_{dd}(r1-r2)=\frac{(e\cdot d)^2}{4\pi\epsilon}\frac{1}{{|r_1-r_2|}^3}$ that diverges when the interparticle distance $r=|r_1-r_2|$ vanishes. To circumvent this singularity we introduce a cut off distance $r_c$ such that we consider the relative motion part of the two particle wave-function $\psi(r)$ when $r<r_c$, while for $r>r_c$, $\psi(r \geq r_c)=1$.

For $r<r_c$, we assume that the lattice potential is homogeneous, so that the two-particle Schroedinger equation in relative coordinates reads
\begin{equation}
(-\frac{\hbar^2}{2\mu}\nabla^2+\frac{e^2d^2}{4\pi\epsilon}\frac{1}{{r}^3}-E)\psi(r)=0
\end{equation}
where $\mu$ denotes the two-exciton reduced mass. Setting $\psi(r) = A\exp{(-u(r))}$ we obtain
\begin{equation}
(-\frac{\hbar^2}{2\mu}(-\nabla^2u +(\nabla u)^2)  +\frac{e^2d^2}{4\pi\epsilon}\frac{1}{{r}^3}-E)\psi(r)=0
\end{equation}
Since $\psi(r)$ is not vanishing, let us neglect $\nabla^2u$ and $E$ for small r, so that when r tends to zero
\begin{equation}
u(r)\simeq 2\sqrt\frac{2 r_0 \mu/m_x}{r}
\end{equation}
where $r_0=\frac{m_x(e\cdot d)^2}{4\pi\epsilon\hbar^2}$ is the typical lengthscale of the dipolar interaction potential in two dimensions. Since $\psi(r_c)=1$, we finally obtain
\begin{align}
\psi(r) &= \exp{(-2\sqrt{2 r_0 \mu/m_x}(\frac{1}{\sqrt{r}}-\frac{1}{\sqrt{r_c}}))} \text{, } r \leq r_c\\
\psi(r) &= 1 \text{, } r \geq r_c
\end{align}

We now express the on-site dipolar interaction, namely
\begin{align}
U_{n}&=\int \mathbf{dr_1dr_2} U_{dd}(\mathbf{r_1-r_2}) |w^n_0(\mathbf{r_1})|^2|w^n_0(\mathbf{r_2})|^2\\
U_{n}&=\int \mathbf{dr_1dr} U_{dd}(r)  |w^n_0(\mathbf{r_1})|^2|w^n_0(\mathbf{r_1-r})|^2
\end{align}
One thus deduces \cite{Markus}
\begin{align}
U_{n}&\simeq\int \mathbf{dr}  \int \mathbf{dr_1} U_{dd}(r)  |w^n_0(\mathbf{r_1})|^4\psi(r)\\
U_{n}&=\int_0^{r_c} \mathbf{dr}  \int \mathbf{dr_1} U_{dd}(r)  |w^n_0(\mathbf{r_1})|^4\exp{(-2\sqrt{2r_0\mu/m_x}(\frac{1}{\sqrt{r}}-\frac{1}{\sqrt{r_c}}))} +\int_{r_c}^{+\infty}\mathbf{dr}  \int \mathbf{dr_1} U_{dd}(r)  |w^n_0(\mathbf{r_1})|^4\\
U_{n}&=v^{sr}_{n} + v^{lr}_{n}
\end{align}
where we have split the on-site interaction $U_n$ in a short-range part $v^{sr}_{n}$ and a long-range part $v^{lr}_{n}$.
\\

The long range contribution reads
\begin{align}
v^{lr}_{n}&=\int_{r_c}^{+\infty}\mathbf{dr}  \int \mathbf{dr_1} U_{dd}(r)  |w^n_0(\mathbf{r_1})|^4\\
v^{lr}_{n}&=\frac{1}{(2\pi)^4}\int_{r_c}^{+\infty}\mathbf{dr } \int \mathbf{dr_1} \int \mathbf{dk_1dk_2} U_{dd}(r)  |\hat{w}^n_0(\mathbf{k_1})|^2|\hat{w}^n_0(\mathbf{k_2})|^2\exp{(i\mathbf{r_1}(\mathbf{k_1}+\mathbf{k_2}))}
\end{align}
where $\hat{w}^n_0$ denotes the Fourier transform of $w^n_0$. Thus we deduce that 
\begin{align}
v^{lr}_{n}&=\frac{1}{(2\pi)^2}\int \mathbf{dk_2} |\hat{w}^n_0(\mathbf{k_2})|^4 \int_{rc}^{+\infty}\mathbf{dr }U_{dd}(r) \\
v^{lr}_{n}&=\frac{V_d}{(2\pi)^2}\int \mathbf{dk_2} |\hat{w}^n_0(\mathbf{k_2})|^4 \int_{-\pi}^{\pi}\int_{rc}^{+\infty} dr  d\theta \cdot 1/r^2\\
v^{lr}_{n}&=\frac{V_d}{2\pi r_c}\int \mathbf{dk_2} |\hat{w}^n_0(\mathbf{k_2})|^4, \hspace{.1cm} \mathrm{where} \hspace{.1cm} V_d=\frac{e^2d^2}{4\pi\epsilon}
\end{align}

On the other hand, for the short range part of the interaction potential we find that
\begin{align}
v^{sr}_{n}&=\int \mathbf{dr_1}  |w^n_0(\mathbf{r_1})|^4 \int_0^{rc} \mathbf{dr} U_{dd}(r)\exp{(-2\sqrt{2r_0\mu/m_x}(\frac{1}{\sqrt{r}}-\frac{1}{\sqrt{r_c}}))}\\
v^{sr}_{n}&=\pi V_d \int \mathbf{dr_1}  |w^n_0(\mathbf{r_1})|^4 \cdot \frac{1}{2 r_0\mu/m_x} (1 + 2\sqrt{\frac{2r_0\mu/m_x}{r_c}})
\end{align}
Summing the long and short-range contributions, we obtain
\begin{align}
U_{n}&=v^{sr}_{n}+ v^{lr}_{n}\\
U_{n}&=V_d  \int \mathbf{dr_1} |w^n_0(\mathbf{r_1})|^4 \cdot \frac{\pi}{2 r_0\mu/m_x} (1 + 2\sqrt{\frac{2r_0\mu/m_x}{r_c}})+\frac{V_d}{2\pi r_c}\int \mathbf{dk_2} |\hat{w}^n_0(\mathbf{k_2})|^4  
\end{align}
Parseval theorem finally leads to
\begin{align}
U_{n}&=V_d  \int \mathbf{dr_1} |w^n_0(\mathbf{r_1})|^4 \cdot (\frac{\pi}{2 r_0\mu/m_x} (1 + 2\sqrt{\frac{2r_0\mu/m_x}{r_c}})+\frac{2\pi}{ r_c})
\end{align}
and since for two excitons, $\mu/m_x=0.5$ we finally deduce
\begin{equation}
U_{n}=\pi V_d  \int \mathbf{dr_1} |w^n_0(\mathbf{r_1})|^4 \cdot ((\frac{1}{ r_0}  + \frac{2}{ \sqrt{r_c r_0}})+\frac{2}{ r_c})
\label{Ud}
\end{equation}

Having computed the Wannier functions (Fig.S8), we evaluate $ \int \mathbf{dr_1} |\hat{w}^n_0(\mathbf{r_1})|^4$ and then calculate Eq (\ref{Ud}) by introducing a cut off distance balancing the short range and long range contributions, $r_c\sim r_0/2$. Furthermore, we verified that the magnitude of $U_{n}$ varies weakly with the exact cut-off distance. Figure S9 presents the amplitude of on-site dipolar repulsions, for every WS of the 800 nm and 400 nm period lattices, panels a) and b) respectively. For the WS where we report the buildup of Mott phases, namely the 7$^\mathrm{th}$ and the 5$^\mathrm{th}$ WS of each lattice, we deduce that the on-site interaction strength is equal to \SI{80}{\micro\eV} and \SI{184}{\micro\eV} respectively. Importantly, we compare these magnitudes to the energy separation between successive WS, \fd{$\Delta E$} (black line in Fig.S9). \fd{Thus, we note that for the 800 nm period lattice $U_{7}$ is slightly smaller than $\Delta E$, and that $U_{5}$ is about $\Delta E$ for the 400 nm lattice}. Finally, as shown in Table 2, we verified that the amplitude of off-site interactions between nearest neighbouring sites is negligible compared to $U$ for both devices, justifying why we have not taken it into account.

\vspace{.5cm}

\centerline{\includegraphics[width=.8\linewidth]{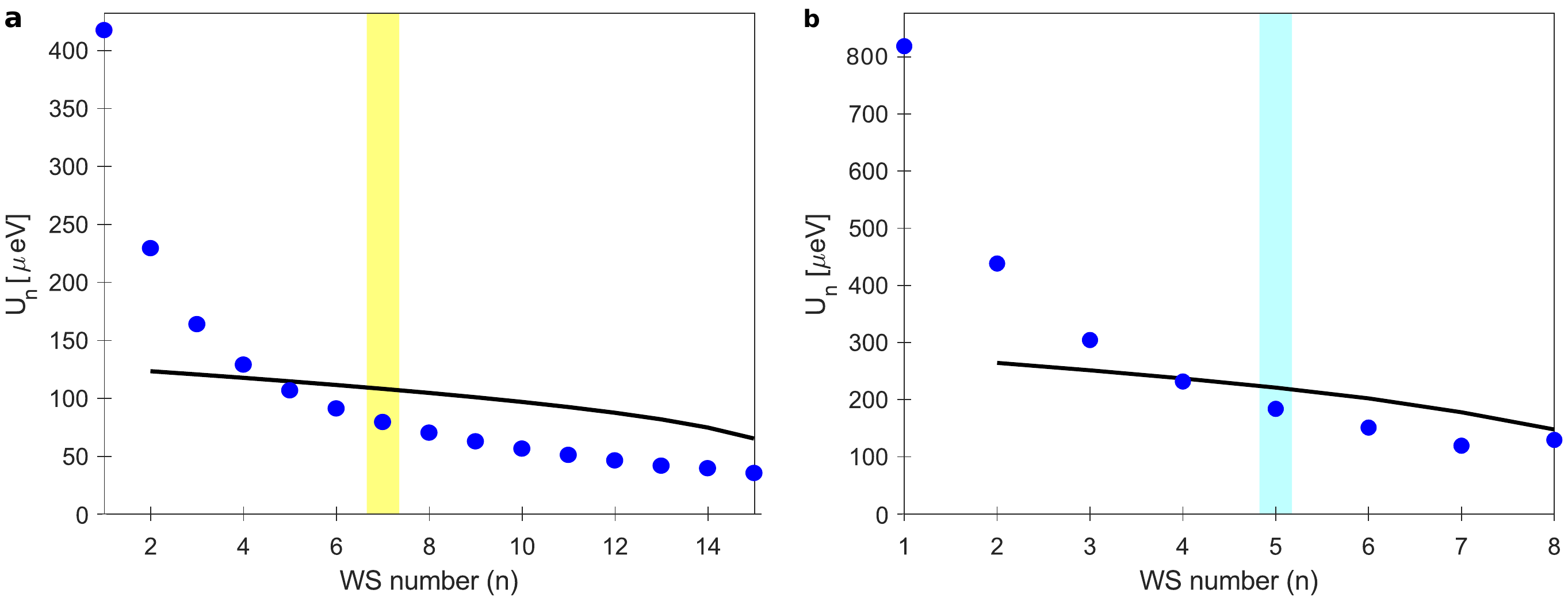}}
\textit{\textbf{Fig. S9}: On-site dipolar interaction coefficient $U_n$ for all Wannier states of the 800 nm period a) and 400 nm period lattice b). We display in blue the magnitude of $U_n$ while the black lines shows the energy difference between successive Wannier states ($\Delta E$). The shaded areas underline the WS where we report Mott phases in each lattice.}

\subsection{Tunnelling amplitudes}

To evaluate the tunnelling amplitudes we use the energies of Bloch bands that directly lead to the hoping coefficients (Eq. \ref{tunnel}). For the nearest neighbour tunnelling strength, one finds \cite{Greiner_PhD}
\begin{align}
t_n&=-\frac{a}{2\pi}\int_{-\pi/a}^{\pi/a}E^{n}_q  e^{(iaq)} dq
\end{align}

Figure S10 displays the magnitudes of $t_n$, in vertical logscale, for all Wannier states of the 800 nm period (a) and 400 nm period lattice (b). Importantly we note that between the 7$^\mathrm{th}$ WS of the 800 nm period lattice and the 5$^\mathrm{th}$ WS of the 400 nm period one the tunnelling amplitude is increased by 6 orders of magnitude.

\vspace{.5cm}
\centerline{\includegraphics[width=.8\linewidth]{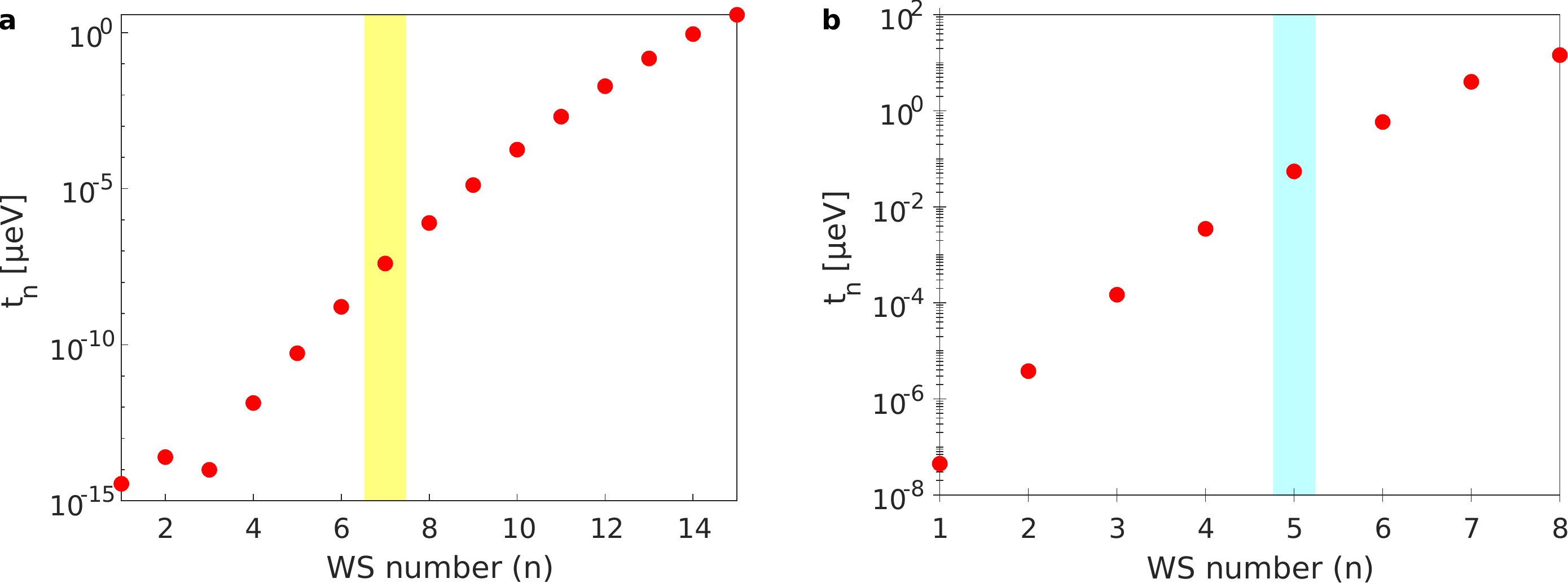}}
\textit{\textbf{Fig. S10}: Hopping coefficient $t_n$ in vertical logscale for all Wannier states of the 800 nm lattice a) and 400 nm lattice b).  The shaded area underline the WS where we report Mott phases in each lattice.}

\vspace{.5cm}

\fd{Finally, we note that possible lattice inhomogeneities do not suffice to strongly alter the excitons tunnelling strength in our experiments. Indeed, Fig.3-4 of the main text underline that the photoluminescence energy does not vary spatially in the MI regime. Accordingly, we deduce that across the lattice the energy of WS varies less than our instrumental resolution (15 $\mu$eV). This reveals that the amplitude of disorder in the lattice is bound to $\Delta_m$=15 $\mu$eV. Following Mott's law of variable range hopping \cite{Mott}, we deduce that the minimum tunnelling strength is then ($t_n\cdot e^{-\Delta_m/k_BT}$), where $t_n$ is the disorder-free tunnelling strength of the $n$-th WS (Fig.S10). For this worst-case situation, we find that during our measurements excitons can tunnel between at least 9 lattice sites for the 5$^{th}$ WS of the 400 nm period lattice. On-site interactions provide therefore the only mechanism to possibly inhibit tunnelling in the measurements shown in Fig.4.a.}

\subsection{Interplay between $U$, $t$ and $\Delta E$ for Mott phases}

Having extracted the magnitudes of $U$ and $t$ for each lattice device, we now compare the ratio ($U/t$) for every WS, since Mott phases are only expected when $(U/t)\gtrsim$20 \cite{Trivedi_2011}. Fig.S11.a-b then show that this condition is largely fulfilled for the WS where we report Mott-like phases in the 400 and 800 nm period lattices. For the former in the 5$^\mathrm{th}$ WS we deduce $(U/t)\sim 3000$ while $(U/t)\sim 10^9$ for the 7$^\mathrm{th}$ WS of the 800 nm period device. 
The first value corresponds to around 2-3 times the strongest interaction regime explored with cold atoms, while the latter one further recalls that in the 800 nm period lattice tunnelling is vanishingly small in the 7$^\mathrm{th}$ WS.

Fig.S11.a-b underlines that for each lattice we find numerous WS such that $(U/t)\gtrsim$20 is verified. One then wonders why Mott phases are observed for the 7$^\mathrm{th}$ and 5$^\mathrm{th}$ WS only for the 800 and 400 nm period lattices respectively. \fd{To address this question, in Fig.S11.c-d we compare the magnitude of ($U+k_BT$) and $\Delta E$ for each WS, thus taking into account that our studies are realised at finite temperature. Remarkably, for our two devices we note that Mott phases are observed in the lowest energy WS such that ($U+k_B T$) does not exceed $\Delta E$. Precisely, for the 800 nm period device we find that  $(U+k_BT)\lesssim \Delta E$, which certainly contributes to stabilising the Mott-like phase with 2 excitons per lattice site. On the other hand, for the 400 nm period device, $(U+k_B T)\simeq \Delta E$. This may explain why Mott phases with 2 excitons per lattice site are not observed.}

\centerline{\includegraphics[width=.7\linewidth]{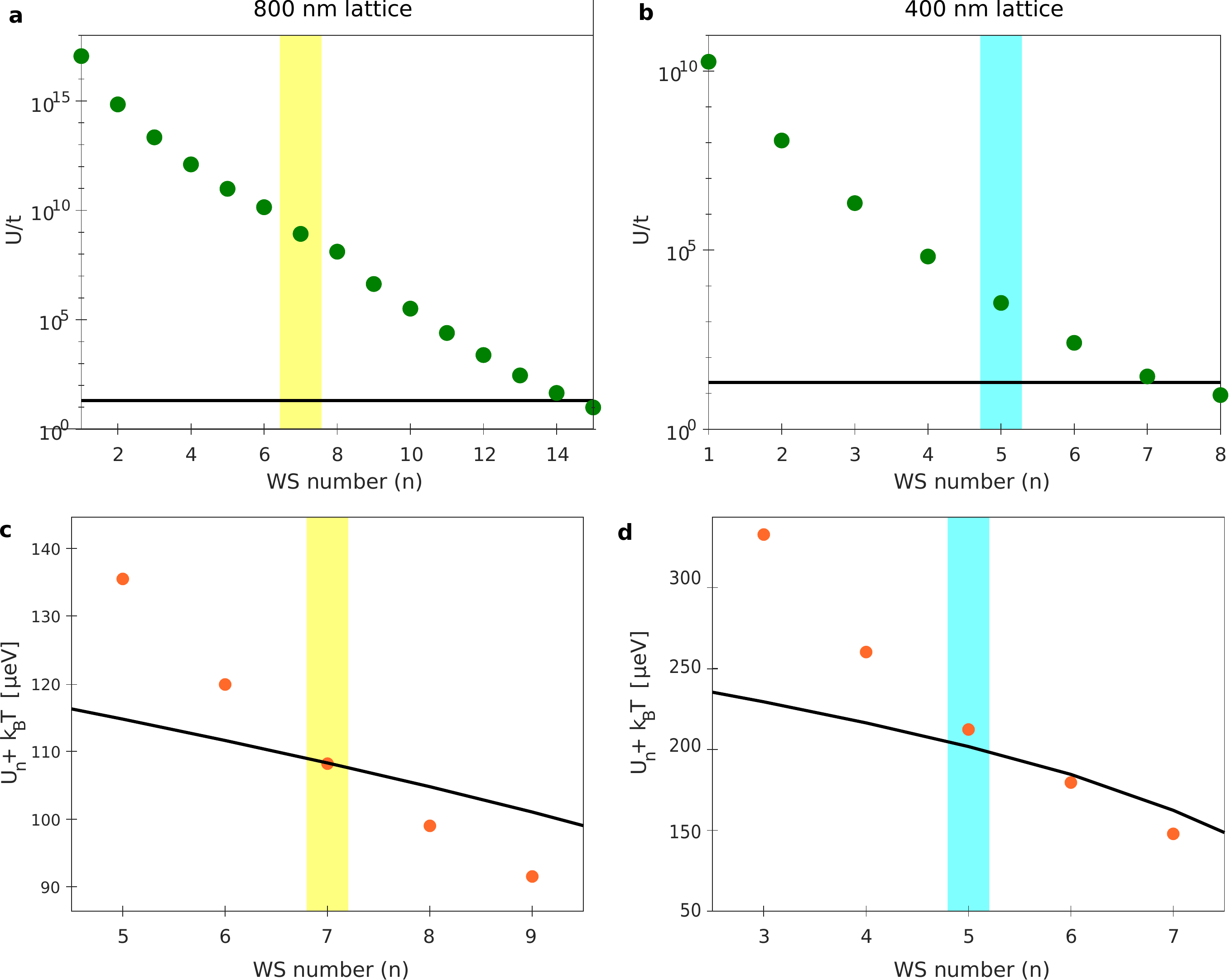}}
\textit{\textbf{Fig. S11}: Ratio ($U/t$) deduced for every WS of the 800 nm (a) and 400 nm (b) period lattice. The solid black line marks the threshold value $(U/t)\sim$20, and the shaded areas highlight the WS where we report Mott phases. For the 800 nm period lattice we then deduce $(U/t)\sim 10^9$ for the 7$^\mathrm{th}$ WS and 3000 for the 5$^\mathrm{th}$ one of the 400 nm period device. c-d) Comparison between the energy separating successive WS (\fd{$\Delta E$} in black) and the strength of on-site interactions to which we add the thermal energy (orange points), for the 800 nm (c) and 400 nm (d) period lattice, at $T=$330 mK.}

\section{Thermal melting of Mott-like phases}

As introduced in the main text, the experiments shown in Figs.2-4 are obtained by averaging 10 measurements lasting typically 30 seconds, so that in the overall 5-minute measurement time 255$\cdot$10$^{6}$ realisations are averaged. This first shows that Mott-like phases are well stabilised in our experiments. Nevertheless, the mean fraction of excitons contributing to them is bound to about 45$\%$. We attribute this limitation to the lowest accessible bath temperature, $T=330$ mK. Indeed, we then have $k_B T/U\sim$ 0.3 for the 800 nm period lattice and $k_B T/U\sim$ 0.15 for the 400 nm period one. These regimes are then in the range where a Mott insulator is expected to start melting into a normal fluid \cite{Bloch_2010,Gerbier_2007,deMArco_2005,Trivedi_2011}. To verify this expectation, we studied for the 800 nm period lattice the variation of the exciton fraction in the $n_X=2$ Mott-like phase as a function of the bath temperature. Figure S12 shows that the occupation fraction of the 7$^\mathrm{th}$ Wannier state $\overline{p(7)}$ is dramatically reduced while the bath temperature is increased, starting from 55$\%$ at 330 mK. On the other hand, the occupation of lower energy states is increased, manifesting that exciton relaxation towards deeper confined levels is more effective. This behaviour was somewhat expected since the energy splitting between WS is around the thermal energy at $T\sim$ 1K. 

To quantify the melting of the $n_X$=2 Mott-like phase we compared $\overline{p(7)}$ to the summed occupation of lower energy WS, $\sum_{i=1,6} \overline{p(i)}$. Indeed, a Mott phase is protected energetically by $U$ \cite{Salomon_2010}, from exciton relaxation that increases the occupation of lower energy states according to Fig.S12.a. Theoretically $\overline{p(7)}/\sum_{i=1,6} \overline{p(i)}$ scales then as $e^{-U/k_BT}$ and Fig.S10.b shows that our measurements follow this behaviour if we set $U=$ 60$\pm$10 $\mu$eV. \fd{Let us note that we reach the same conclusion by computing $\overline{p(7)}/\sum_{i\neq7} \overline{p(i)}$.} Thus, we confirm the theoretical magnitude for on-site dipolar repulsions  (\SI{80}{\micro\eV}). Moreover, we verify that the Mott-like phase melts very rapidly since our lowest bath temperature is relatively  high \cite{Bloch_2010}. 

\vspace{.5cm}

 \centerline{ \includegraphics[width=.75\linewidth]{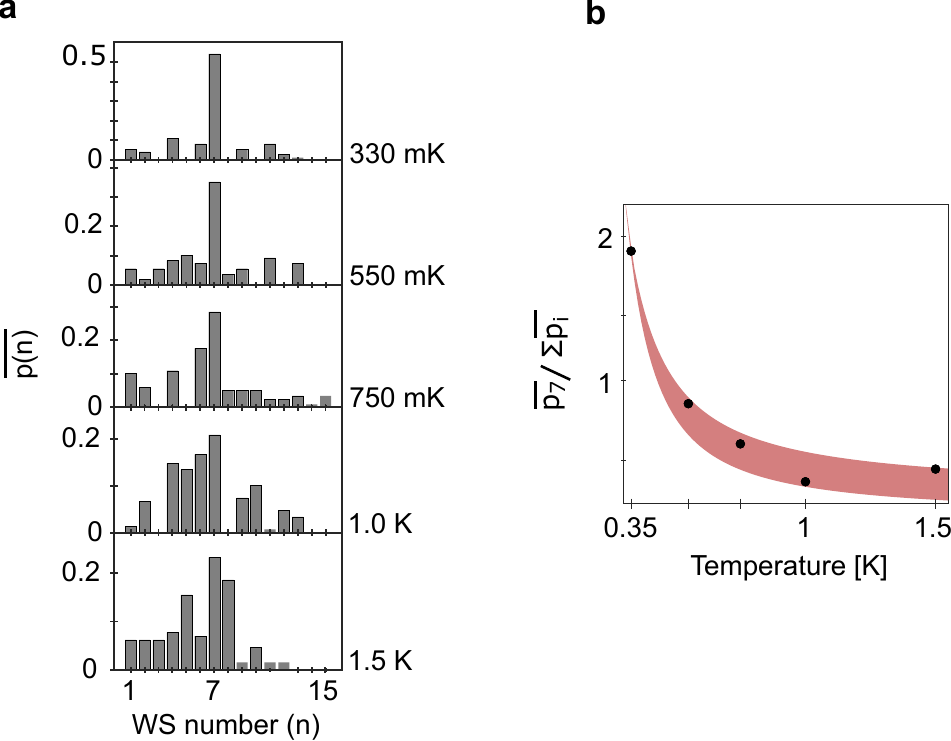}}
  \textit{\textbf{Fig. S12:} a) Occupation fraction of the 15 Wannier states of the 800 nm period lattice as a function of the bath temperature, from 330 mK to 1.5 K from top to bottom. In every case, $P$=3 nW so that the system is initially prepared in the $n_X=2$ Mott-like phase. b) Ratio between the occupation fraction of the 7$^\mathrm{th}$ WS $\overline{p(7)}$ and the summed occupation of lower energy states $\sum_{i=1-6} \overline{p(i)}$. The red shaded area marks an exponential decrease $e^{-U/k_BT}$ setting $U=60\pm10 \mu$eV.}

\newpage

\section{Summary of main physical parameters}

Below, in Table 2 we summarise most relevant physical parameters that are necessary to quantify our experimental findings.

\begin{center}
   \begin{tabular}{  | l | c | c | }
     \hline
          Lattice period & \SI{800}{\nano\metre}  & \SI{400}{\nano\metre}  \\ \hline
            \multicolumn{3}{|c|}{Electrostatic parameters} \\  \hline
Exciton dipole (e$\cdot$d)&\multicolumn{2}{c|}{ 575.6 Debye}\\  \hline
Voltage applied between surface and ground electrodes (V)& -2.0 V & -1.4 V \\     \hline
Energy of the photoluminescence of direct exciton ($E_{DX}$)&\multicolumn{2}{c|}{1577 meV (\SI{787}{\nano\metre})}\\  \hline
Energy of the photoluminescence of indirect exciton ($E_{IX}$)& 1527 meV & 1520 meV \\     \hline
            \multicolumn{3}{|c|}{Lattice potential in the DQW plane} \\  \hline
Lattice potential depth ($V_0$)&\multicolumn{2}{c|}{ 1.5 meV}\\  \hline
In plane electric field maximum amplitude ($|E_r|$)& \SI{0.1}{\volt\per\micro\metre}& \SI{0.14}{\volt\per\micro\metre} \\     \hline
Mean value of the z electric field componant ($<E_z>$)& \SI{3.08}{\volt\per\micro\metre} &  \SI{3.86}{\volt\per\micro\metre}\\     \hline
$|E_r|/<E_z>$& 3.31\% & 3.36\%\\     \hline
            \multicolumn{3}{|c|}{ Bose Hubbard model parameters} \\  \hline
						Lattice connectivity (z)&\multicolumn{2}{c|}{ 4}\\  \hline
Band of interest& 7 & 5 \\     \hline
Recoil energy ($E_r$)& \SI{2.7}{\micro\eV} & \SI{10.6}{\micro\eV} \\     \hline
$s=V_0/E_r$& 565 & 142\\     \hline
Calculated tunnel coefficient (t)& \SI{4e-8}{\micro\eV}  & \SI{0.06}{\micro\eV}  \\     \hline
Tunneling time ($\hbar/(zt)$)& 4 ms & 3 ns\\     \hline
Calculated on-site dipolar interaction coefficient (U)& \SI{80}{\micro\eV} & \SI{184}{\micro\eV} \\     \hline
Thermal energy at $T$=330 mK ($k_BT$)& \multicolumn{2}{c|}{ \SI{28.5}{\micro\eV} } \\     \hline
$U/k_BT$& 2.8 & 6.4 \\     \hline
Calculated off-site dipolar interaction coefficient ($V_1$)& \SI{32}{\nano\eV}  & \SI{250}{\nano\eV}  \\     \hline
Typical lengthscale of the dipolar interaction in two dimensions ($r_0$)&\multicolumn{2}{c|}{ \SI{50}{\nano\metre}}\\  \hline
             \multicolumn{3}{|c|}{Material properties} \\  \hline
Gallium arsenite permittivity ($\epsilon$)&   \multicolumn{2}{c|}{ 12.5 $\epsilon_0$}\\  \hline
Exciton effective mass ($m_x$) & \multicolumn{2}{c|}{ 0.22 $m_e$}\\  \hline
2 Exciton reduced mass ($\mu_x$) & \multicolumn{2}{c|}{ 0.5 $m_x$}\\  \hline
             \multicolumn{3}{|c|}{Fundamental constants} \\  \hline
Elementary charge (e)&   \multicolumn{2}{c|}{\SI{1.6e-19}{\coulomb} }\\  \hline
Boltzmann constant ($k_B$)&   \multicolumn{2}{c|}{\SI{1.38E-23}{\joule\per\kelvin}}\\  \hline
Planck constant ($\hbar$)&   \multicolumn{2}{c|}{\SI{1.05E-34}{\joule\second}}\\  \hline
Electron mass ($m_e$)&\multicolumn{2}{c|}{\SI{9.1e-31}{\kilo\gram}}\\  \hline
Vacuum permittivity ($\epsilon_0$)&   \multicolumn{2}{c|}{\SI{8.854e-12}{\farad\per\metre}}\\  \hline
\end{tabular}
 \end{center}
 \centerline{ \textit{\textbf{Table 2:} Summary of main physical quantities}}

\end{document}